\documentclass[11pt,a4paper]{article}
\usepackage{amssymb,amsmath,amsfonts,amsthm,amstext,amscd,array}
\usepackage{mathrsfs}
\usepackage{hyperref}
\usepackage{pdfsync}
\usepackage{bbm}
\usepackage{bm}
\usepackage[arrow,matrix,curve]{xy}
\usepackage{bbding}
\usepackage{wasysym}

\usepackage{epsf}
\usepackage{epsfig}
\usepackage{wrapfig}

\input{epsf}

\def\nn{\nonumber}

\newcommand{\Z}{\zed}
\newcommand{\N}{\nat}

\newcommand{\mbf}[1]{{\boldsymbol {#1} }}
\def\ii{{\,{\rm i}\,}}
\def\dd{{\rm d}}

\def\sfs{{\sf s}}

\def\Diff{{\sf Diff}}

\def\Tw{\widehat{T}{}}

\def\mcH{{\mathcal H}}

\def\mcE{{\mathcal E}}

\newcommand{\CCC}{\mathscr{C}}

\newcommand{\CCP}{\mathscr{P}}
\newcommand{\CCF}{\mathscr{F}}

\newcommand{\CCH}{\mathscr{H}}
\newcommand{\CP}{\mathcal{P}}

\newcommand{\CJ}{\mathcal{J}}
\newcommand{\CI}{\mathcal{I}}

\newcommand{\eq}{\begin{equation}}
\newcommand{\eqend}{\end{equation}}
\newcommand{\eqa}{\begin{eqnarray}}
\newcommand{\nonueqa}{\begin{eqnarray*}}
\newcommand{\eqaend}{\end{eqnarray}}
\newcommand{\nonueqaend}{\end{eqnarray*}}

\newcommand{\bma}[1]{\begin{array}{#1}}
\newcommand{\ema}{\end{array}}
\newcommand{\bc}{\begin{center}}
\newcommand{\ec}{\end{center}}

\newcommand{\R}{\real}

\renewcommand{\thefootnote}{\fnsymbol{footnote}}
\newcommand{\newsection}{\setcounter{equation}{0}\section}

\newcommand{\zed}{{\mathbb Z}} 
\newcommand{\nat}{{\mathbb N}} 
\newcommand{\real}{{\mathbb R}} 

\def\alg{{\mathcal A}}
\def\balg{{\mathcal B}}

\newif\ifold             \oldtrue

\def\nn{\nonumber}

\def\e{{\,\rm e}\,}

\def\be{\begin{equation}}
\def\ee{\end{equation}}
\def\bea{\begin{eqnarray}}
\def\eea{\end{eqnarray}}
\def\bd{\begin{displaymath}}
\def\ed{\end{displaymath}}

\newcommand{\beq}{\begin{eqnarray}}
\newcommand{\eeq}{\end{eqnarray}}

\makeatletter
\newdimen\normalarrayskip              
\newdimen\minarrayskip                 
\normalarrayskip\baselineskip
\minarrayskip\jot
\newif\ifold             \oldtrue            
\def\arraymode{\ifold\relax\else\displaystyle\fi} 
\def\@arrayskip{\ifold\baselineskip\z@\lineskip\z@
     \else
     \baselineskip\minarrayskip\lineskip2\minarrayskip\fi}
\def\@arrayclassz{\ifcase \@lastchclass \@acolampacol \or
\@ampacol \or \or \or \@addamp \or
   \@acolampacol \or \@firstampfalse \@acol \fi
\edef\@preamble{\@preamble
  \ifcase \@chnum
     \hfil$\relax\arraymode\@sharp$\hfil
     \or $\relax\arraymode\@sharp$\hfil
     \or \hfil$\relax\arraymode\@sharp$\fi}}
\def\@array[#1]#2{\setbox\@arstrutbox=\hbox{\vrule
     height\arraystretch \ht\strutbox
     depth\arraystretch \dp\strutbox
     width\z@}\@mkpream{#2}\edef\@preamble{\halign \noexpand\@halignto
\bgroup \tabskip\z@ \@arstrut \@preamble \tabskip\z@ \cr}%
\let\@startpbox\@@startpbox \let\@endpbox\@@endpbox
  \if #1t\vtop \else \if#1b\vbox \else \vcenter \fi\fi
  \bgroup \let\par\relax
  \let\@sharp##\let\protect\relax
  \@arrayskip\@preamble}
\makeatother

\newcommand{\p}{\partial}

\def\FF{{\cal F}}

\def\be{\beta}

\theoremstyle{definition}

\def\cirp{\mathop{\bar\circ}}

\usepackage{hyperref}
\hypersetup{colorlinks,%
citecolor=black,%
filecolor=black,%
linkcolor=black,%
urlcolor=black}

\begin{document}

\begin{titlepage}

\begin{flushright}
\small
\baselineskip=12pt
MPP--2018--28\\
EMPG--18--05
\end{flushright}
\normalsize

\begin{center}

\vspace{1cm}

\baselineskip=24pt

{\Large\bf Symplectic realisation of electric charge\\ in fields of monopole distributions}

\baselineskip=14pt

\vspace{1cm}

{\bf Vladislav G. Kupriyanov}${}^{1}$ \ and \ {\bf Richard
  J. Szabo}${}^{2}$
\\[5mm]
\noindent ${}^1$ {\it Max-Planck-Institut f\"ur Physik,
  Werner-Heisenberg-Institut\\ F\"ohringer Ring 6, 80805 M\"unchen, Germany
}
\\ and {\it CMCC-Universidade de Federal do ABC, Santo Andr\'e, SP, 
Brazil}\\ and {\it 
Tomsk State University, Tomsk, Russia}
\\
Email: \ {\tt
    vladislav.kupriyanov@gmail.com}
\\[3mm]
\noindent  ${}^2$ {\it Department of Mathematics, Heriot-Watt University\\ Colin Maclaurin Building,
  Riccarton, Edinburgh EH14 4AS, U.K.}\\ and {\it Maxwell Institute for
Mathematical Sciences, Edinburgh, U.K.} \\ and {\it The Higgs Centre
for Theoretical Physics, Edinburgh, U.K.}\\
Email: \ {\tt R.J.Szabo@hw.ac.uk}
\\[30mm]

\end{center}

\begin{abstract}
\baselineskip=12pt
\noindent
We construct a symplectic realisation of the twisted Poisson structure
on the phase space of an electric charge in the background of an
arbitrary smooth magnetic monopole density in three dimensions. We use
the extended phase space variables to study the classical and quantum
dynamics of charged particles in arbitrary magnetic fields by
constructing a suitable Hamiltonian that reproduces the Lorentz force
law for the physical degrees of freedom. In the source-free case the
auxiliary variables can be eliminated via Hamiltonian reduction, while
for non-zero monopole densities they are necessary for a consistent
formulation and are related to the extra degrees of freedom usually
required in the Hamiltonian description of dissipative systems. We
obtain new perspectives on the dynamics of dyons and motion in the
field of a Dirac monopole, which can be formulated without Dirac
strings. We compare our associative phase space formalism with the
approach based on nonassociative quantum mechanics, reproducing
extended versions of the characteristic translation group
three-cocycles and minimal momentum space volumes, and prove that the
two approaches are formally equivalent. We also comment on the
implications of our symplectic realisation in the dual framework of
non-geometric string theory and double field theory.
\end{abstract}

\end{titlepage}
\setcounter{page}{2}

\newpage

{\baselineskip=12pt
\tableofcontents
}

\bigskip

\renewcommand{\thefootnote}{\arabic{footnote}}
\setcounter{footnote}{0}

\newsection{Introduction and summary\label{sec:intro}}

Despite their elusiveness to experimental observation, magnetic monopoles
have been of wide{-}spread theoretical interest in various areas of physics for many years due
to their novel conceptual and mathematical implications. In particular, the
quantum mechanics of an electric charge coupled to a magnetic monopole
density exhibits a variety of interesting geometric and algebraic
features. For the standard example of motion in the field of a Dirac
monopole, the charged particle wavefunction can be regarded as a
section of a non-trivial line bundle associated to the Hopf
fibration~\cite{Wu1976} which provides a topological explanation for
Dirac charge quantisation~\cite{Dirac1931,Dirac1948} and
formulates the quantum dynamics of the particle without using the
unphysical Dirac string singularities that usually arise due to the
absence of a globally defined magnetic vector potential for the monopole
field. 

In this paper we are predominantly interested in smooth distributions of magnetic charge, for which vector potentials do not exist even locally and the classical dynamics of the canonical phase space coordinates of the particle are described by a necessarily nonassociative twisted Poisson algebra. These systems have been of interest recently as magnetic analogues of certain flux models in non-geometric string theory and double field theory, see e.g.~\cite{Lust2012,BL,Blumenhagen2013,Mylonas2014,Lust:2017bgx} and references therein.
In these instances the corresponding quantum theory cannot be
formulated in the usual framework of canonical quantisation by
operators acting on a separable Hilbert space. Two formulations have
thus far been proposed to handle nonassociative quantum mechanics in
this setting, each with its own limitations. Deformation quantisation
by explicit construction of nonassociative phase space star products was
originally developed by~\cite{Mylonas2012}, and subsequently treated
in~\cite{BL,Mylonas2013,Kupriyanov2015}; however, beyond the case of
constant magnetic charge density, this procedure does not yield a
quantisation of the classical dynamical system because the Planck constant $\hbar$
appears as a formal expansion parameter, and the result is a
deformation over an algebra of formal power series rather than with a
complex parameter. On the other hand, the approach
of~\cite{Bojowald2014,Bojowald2015} is based on developing algebraic
properties of quantum moments from the assumption that the underlying
twisted Poisson algebra is a Malcev algebra; however, even in the
simplest case of constant monopole density, the alternative property
used there cannot be realised in general and this formalism is restricted to observables which are linear in the kinematical momenta.

The purpose of the present paper is to develop a new approach to the
classical and quantum dynamics of electric charges in monopole
distributions by generalising the technique of `symplectic
realisation' from Poisson
geometry~\cite{Weinstein-local,Karasev,Weinstein-groupoid,Weinstein}. Symplectic
realisation is a useful mathematical tool for quantisation, because it
embeds arbitrary Poisson manifolds into the framework of geometric
quantisation or other standard quantisation methods based on
symplectic structures. In the following we construct a symplectic
realisation of the twisted Poisson structure corresponding to the
algebra of covariant momenta of a charged particle in the background
of an arbitrary monopole field. This construction doubles the original
phase space coordinates by introducing a set of auxiliary degrees of
freedom. The resulting associative extended algebra of Poisson brackets is then used to construct a Hamiltonian description of a charged particle interacting with a distribution of magnetic monopoles, by an appropriate choice of Hamiltonian on the extended phase space that leads to the Lorentz force equation for the physical coordinates. In the case of a magnetic field with no monopole sources, one can eliminate the auxiliary variables by Hamiltonian reduction and thus recover the standard Hamiltonian formulation of a  charged particle in a divergenceless magnetic field. However, in the presence of magnetic monopoles, the auxiliary degrees of freedom are necessary for a consistent Hamiltonian description. In the particular example of a spherically symmetric magnetic field sourced by a constant monopole density, we show that Hamiltonian reduction results either in free particle motion or in the absence of propagating degrees of freedom altogether. We demonstrate that the necessary presence of auxiliary degrees of freedom in this case is related to the fact that the motion of a charged particle can be effectively described as motion in the field of a single Dirac monopole with some frictional forces~\cite{BL}, and normally the consistent Hamiltonian description of a system with friction requires the introdution of additional degrees of freedom representing a reservoir.

With this set up, quantisation then proceeds along the usual lines using canonical methods by
constructing a suitable Hilbert space on which the quantum
Hamiltonian operator acts, and studying the Schr\"odinger equation. Our
formalism mimicks the standard quantisation schemes which assume that the
phase space is topologically trivial
and that the magnetic field has a globally defined vector potential.
In standard approaches this is not the case even for the Dirac monopole field. Our formalism
also provides a new perspective on the quantisation of a charged
particle in the field of a Dirac monopole, in that our extended vector
potentials are constructed without the usual Dirac string
singularities~\cite{Dirac1931,Dirac1948}. In this respect our approach
of employing extended coordinates is reminescent of old
approaches to the description of electrodynamics
with electric and magnetic sources in terms of two vector
potentials~\cite{Cabibbo1962}, which also avoids the pathologies
associated to the unphysical Dirac string; however,
this latter formulation necessitates Dirac charge quantisation for
consistency, whereas our approach does not. Our constructions analogously have a natural extension to settings which respect
electromagnetic duality, and when applied to the dynamics of dyons, our formalism circumvents the usual problems with defining electromagnetically
dual vector potentials. Gauge theory versions of (ordinary associative) phase space doubling were introduced by~\cite{Faddeev1984,Faddeev1986} for dealing with Schwinger terms viewed as cocycles, and more generally by~\cite{Batalin1986,Batalin1989} for dealing with second class constraints; see~\cite{Feinberg1991} for an application of this formalism to the superparticle and to the Proca Lagrangian.

We will demonstrate that the symplectic realisation which we develop is
equivalent to the framework of nonassociative phase space quantum
mechanics in terms of star products of states and composition products of
observables~\cite{Mylonas2013}, see~\cite{SzaboISQS} for a review in the setting of the present paper. On the other hand, our framework
avoids the problems with constructing star products for spatially varying
monopole densities. Our formalism should reproduce the quantum
fluctuations which compute the quantum evolution of basic dynamical
variables from~\cite{Bojowald2014,Bojowald2015}, which demonstrate
that the nonassociative dynamics generically exhibit modifications of
the classical Lorentz force law (but we do not check this in detail). In our formalism based on an extended
phase space we are able to reproduce the novel predictions of
nonassociative quantum mechanics within an associative approach, such as an extended realisation of
the three-cocycle of the translation group which
obstructs a projective representation on the charged particle
wavefunctions, and also of the minimal uncertainty volumes due to
non-vanishing associators. The latter are particularly interesting in
the string theory dual models where they imply a coarse-graining of
spacetime~\cite{Blumenhagen2010,Lust2010,Mylonas2013}. Indeed, our framework is
analogous to the locally `non-geometric'
backgrounds in string theory, wherein there are no local expressions
for the geometry and the background fields require the
extended space of double field theory for their proper definition. In particular, as discussed
by e.g.~\cite{Lust:2017bgx}, the uniform magnetic charge
density is the magnetic analogue of the locally non-geometric $R$-flux
background of string theory in three dimensions. Our approach bears
certain qualitative similarities to double field theory, such as an
underlying $O(3,3)\times O(3,3)$ symmetry of the dynamics on the extended
phase space, but also certain important differences that we
discuss in the following.

The outline of the remainder of this paper is as follows. In
Section~\ref{sec:monopolealg} we introduce our symplectic realisation
of the twisted Poisson algebra governing the kinematics of an electric
charge in a generic magnetic background; our approach generalises the
standard symplectic realisations of Poisson structures whose technical
details we briefly describe in
Appendix~\ref{app:symplecticPoisson}. In Section~\ref{sec:classdyn} we
demonstrate how to construct a suitable Hamiltonian on the extended
phase space which reproduces the classical Lorentz force law for the
physical degrees of freedom, while in Section~\ref{sec:Hamred} we analyse in
detail the problem of classical Hamiltonian reduction, or
polarisation, of the extended dynamics. In Section~\ref{sec:Sduality}
we extend our formalism in a way which respects electromagnetic
duality and describe the ensuing dynamics of dyons within our
symplectic realisation. In Section~\ref{sec:integrability} we address
the problem of explicitly integrating the equations of motion for a
charged particle in a magnetic monopole background; for spherically
symmetric magnetic fields sourced by constant monopole densities we
relate the classical dynamics on extended phase space to that of
dissipative systems, which is reviewed briefly in
Appendix~\ref{app:dho}, and we show that the equations of motion are
integrable in the case of axial magnetic fields, obtaining the
explicit solution which is analogous to motion in the source-free case. In
Section~\ref{sec:quantumdynamics} we consider the quantum dynamics in
the symplectic realisation, demonstrating how the extended formalism reproduces the expected
results in source-free cases, and exploiting the connections to
quantum dissipation discussed in Appendix~\ref{app:dho}. In
Section~\ref{sec:magnetictranslations} we demonstrate how our extended
associative formulation captures features of nonassocative quantum
mechanics, and in Section~\ref{sec:NAQM} we show that the symplectic
realisation is formally equivalent to the phase space formulation of
nonassociative quantum mechanics in terms of star products and
composition products. A final Appendix~\ref{app:nongeometric} compares
the extended phase space of the symplectic realisation to the phase
space of locally non-geometric closed strings and double field theory, and
also presents potential applications of our formalism in non-geometric M-theory.

\newsection{Symplectic realisation of the magnetic monopole
  algebra\label{sec:monopolealg}}

The phase space $\CP$ of a nonrelativistic point particle with electric charge $e$ and
mass $m$ in a magnetic field $\vec B(\vec x\,)$ in three dimensions is
parameterised by position coordinates $\vec x\in\real^3$ and the
kinematical momentum variables $\vec{\bar\pi}=m\,\dot{\vec x}$, where an
overdot denotes a time derivative. The coordinate algebra is defined by the brackets
\bea
\{x^i,x^j\}= 0 \ , \quad \{x^i,\bar\pi_j\}=\delta^i{}_j \qquad \mbox{and} \qquad \{\bar\pi_i,\bar\pi_j\}=e\, \varepsilon_{ijk} \, B^k(\vec x\,) \ ,\label{a1}
\eea
where throughout we set the speed of light to $c=1$.
These brackets are equivalent to an almost symplectic structure on $\CP$, i.e. a
non-degenerate two-form on phase space given by
\bea\label{eq:omegamonopole}
\omega = \mbox{$\frac e2$}\, \varepsilon_{ijk}\, B^k(\vec x\,)\ \dd x^i\wedge \dd x^j + \dd\bar\pi_i\wedge \dd x^i \ ,
\eea
which is not generally closed:
The Jacobiator amongst the momenta $\bar\pi_i$ is given by
\bea
\{\bar\pi_i,\bar\pi_j,\bar\pi_k\}:= \mbox{$\frac13$}\, \{\bar\pi_i,\{\bar\pi_j,\bar\pi_k\}\}+\mbox{cyclic} =e\,\varepsilon_{ijk}\,\vec \nabla\cdot\vec B \ .\label{a2}
\eea
Thus the brackets \eqref{a1} generically define a twisted Poisson
structure on the six-dimensional phase space, with twisting three-form
$\tau=\dd\omega= \frac{e}{3!} \,\varepsilon_{ijk}\,(\vec
\nabla\cdot\vec B\,) \, \dd x^i\wedge\dd x^j\wedge\dd x^k$ on the
configuration space.\footnote{See Appendix~\ref{app:symplecticPoisson}
  for relevant background on Poisson geometry, and in particular for a
  brief account of the general theory of symplectic realisations that
  we use below.}

In classical Maxwell theory one has $\vec\nabla\cdot\vec B=0$, so that the Jacobiator (\ref{a2}) vanishes and the
algebra (\ref{a1}) is associative. In this case the magnetic field can
be written as $\vec B=\vec
\nabla\times \vec a$ for a globally defined smooth vector potential $\vec
a(\vec x\,)$ on $\real^3$, and the Poisson algebra can be represented by transforming the symplectic two-form \eqref{eq:omegamonopole} to the standard symplectic structure $\omega= \dd p_i\wedge\dd x^i$ for the canonically conjugate position and momentum coordinates $\vec x$ and $\vec
p=\vec{\bar\pi}+e\,\vec a$.

For the spherically symmetric field
\bea
\vec B_{\rm D}(\vec x\,) = g\, \frac{\vec x}{|\vec x\,|^3}
\label{eq:BDirac}\eea
of a Dirac monopole at the origin with magnetic charge $g$, the
algebra is associative at every point away from the location of the
monopole. In this case one can excise the origin, where $\vec B_{\rm
  D}$ is singular, and locally define a corresponding vector
potential~\cite{Zwanziger1968}
\bea\label{eq:aDirac}
\vec a_{\rm D}(\vec x\,) = \frac g{|\vec x\,|} \, \frac{\vec
  x\times\vec n}{|\vec x\,|-\vec x\cdot\vec n}
\eea
at every position $\vec x$ of the
configuration space $\real^3\setminus\{\vec0\,\}$, which has a smaller
domain of regularity obtained by removing a Dirac string singularity~\cite{Dirac1931,Dirac1948} along the line
in the direction of the fixed unit vector $\vec n$ from the origin to
infinity in $\real^3$. The exclusion of the origin from the
configuration space implies that the charged particle and the monopole
cannot simultaneously occupy the same point in space.

In the present paper we are primarily interested in smooth distributions of
magnetic poles, where $\vec \nabla\cdot \vec B\neq0$ on a connected
dense open subset of $\real^3$. In this case there is no associated local
vector potential and the algebra (\ref{a1}) is nonassociative
on the entire configuration space $\real^3$. We call it the {magnetic monopole algebra}.

In this section we will construct a symplectic embedding of the
magnetic monopole algebra. For this, let us introduce an
extended $12$-dimensional phase space $\CCP$ with position coordinates
$x^I=(x^i,\tilde x^i)$ and momentum coordinates $p_I=(p_i,\tilde p_i)$, where
$i=1,2,3$ and $I=1,\dots,6$, which have the canonical Poisson brackets
$\{x^I,x^J\}=\{p_I,p_J\}=0$ and $\{x^I,p_J\}=\delta^I{}_J$. We
define the ``covariant'' momenta associated to the magnetic field
$\vec B(\vec x\,)$ by
\bea
 \pi_I=\left( \pi_i,\tilde \pi_i\right)=p_I-e\,A_I(x^I)\ ,\label{a3}
\eea
with the corresponding magnetic Poisson brackets
\bea\label{eq:piIpijcurl}
\{\pi_I,\pi_J\} = e\, F_{IJ}(x^I) \qquad \mbox{with}  \quad
F_{IJ}= \partial_IA_J-\partial_JA_I \ ,
\eea
where $\partial_I=\frac\partial{\partial x^I}$. We choose the magnetic vector potential $A_I=( A_i,{\tilde A}_i)$ as 
\bea\label{eq:extendedA}
\vec A(x^I)=-\mbox{$\frac12$} \,\vec{\tilde
  x}\times\vec B(\vec x\,) \qquad \mbox{and} \qquad \vec{\tilde
  A}(x^I)= \vec 0\ .\label{a4}
\eea
It has the property
\bea\label{eq:vecpotextended}
\vec{\tilde\nabla}\times\vec A = \vec B \ ,
\eea
where
$\vec{\tilde\nabla}$ are the directional derivatives along the
extended configuration space directions $\vec{\tilde x}$, which we
denote by
$\tilde\partial{}_i=\frac\partial{\partial\tilde x{}^i}$. The non-vanishing Poisson brackets read
\bea
\{x^i,\pi_j\}&=&\{\tilde x^i,\tilde\pi_j\} \ \ = \ \ \delta^i{}_j \ , \nn\\[4pt]
\{\pi_i,\pi_j\}&=& \mbox{$\frac e2$} \,\big(\varepsilon_{ijk}\, \p_l B^k(\vec x\,)-\varepsilon_{ijl}\, \p_k B^k(\vec x\,)\big)\, \tilde x^l\ , \nn\\[4pt]
\{\pi_i,\tilde\pi_j\}&=&\{\tilde\pi_i,\pi_j\} \ \ = \ \ \mbox{$\frac e2$}\, \varepsilon_{ijk}\, B^k(\vec x\,) \ .\label{a5}
\eea
The algebra (\ref{a5}) defines a symplectic structure, and (\ref{a3})
expresses $\pi_I$ in terms of the Darboux coordinates $x^I$ and
$p_I$. 

To relate (\ref{a5}) with the magnetic monopole algebra (\ref{a1}) we introduce
\bea
\vec{\bar\pi}= \vec\pi+\vec{\tilde\pi} \ .
\eea
The vanishing brackets are
\bea\label{PB0}
\{x^i,x^j\}=\{x^i,\tilde x^j\}=\{\tilde x^i,\tilde x^j\}=\{ x^i,\tilde
\pi_j\}=\{\tilde \pi_i,\tilde \pi_j\}=0 \ ,
\eea
while the non-vanishing brackets are given by
\bea
\{x^i,\bar\pi_j\}&=&\{\tilde x^i,\bar\pi_j\}\ \ =\ \ \{\tilde
x^i,\tilde\pi_j \}\ \ = \ \ \delta^i{}_j \ , \nn\\[4pt]
\{\bar\pi_i,\bar\pi_j\}&=&e \, \varepsilon_{ijk} \, B^k(\vec
x\,)+\mbox{$\frac e2$} \,\big(\varepsilon_{ijk}\, \p_l B^k(\vec
x\,)-\varepsilon_{ijl}\, \p_k B^k(\vec x\,)\big)\, \tilde x^l\ , \nn\\[4pt]
\{\bar\pi_i,\tilde\pi_j\}&=&\{\tilde\pi_i,\bar\pi_j\} \ \ = \ \ 
\mbox{$\frac e2$}\, \varepsilon_{ijk}\, B^k(\vec x\,) \ . \label{PB}
\eea
The symplectic
two-form on the extended phase space $\CCP$ corresponding to the Poisson brackets (\ref{PB}) is given by
\bea\label{eq:Omegaextended}
\Omega &=& \mbox{$\frac e2$}\, \big[\,\varepsilon_{ijk}\, B^k(\vec x\,) + \mbox{$\frac12$}\, \big(\varepsilon_{ijk}\, \p_l B^k(\vec
x\,)-\varepsilon_{ijl}\, \p_k B^k(\vec x\,)\big)\, \tilde x^l\,\big]\ \dd x^i\wedge\dd x^j+ \mbox{$\frac e2$}\, \varepsilon_{ijk}\, B^k(\vec x\,)\ \dd x^i\wedge\dd\tilde x^j \nn\\
&& +\, \dd\bar\pi_i\wedge\dd x^i + \dd\bar\pi_i\wedge\dd\tilde x^i + \dd\tilde\pi_i\wedge\dd\tilde x^i \ .
\eea
There is a natural projection ${\sf p}:\CCP\to\CP$ with
${\sf p}(\vec x,\vec{\tilde
  x},\vec{\bar\pi},\vec{\tilde\pi}\,)=(\vec x,\vec{\bar\pi}\,)$, under
which our original phase
space $\CP$ can be embedded into $\CCP$ as the zero section $
{\sf s}_0:\CP\to \CCP$ with ${\sf s}_0(\vec x,\vec{\bar\pi}\,)=(\vec x,\vec0,\vec{\bar\pi},\vec0\,)$. Then the pullback of
\eqref{eq:Omegaextended} to the constraint surface $\CCC_0\subset\CCP$
defined by $\vec{\tilde
  x}=\vec{\tilde\pi}=\vec0$ 
coincides exactly with the almost symplectic structure
\eqref{eq:omegamonopole}, i.e. $
{\sf s}_0^*\,\Omega=\Omega|_{\CCC_0}= \omega$. This therefore determines a symplectic realisation
of the twisted Poisson structure on $\CP$ (see Appendix~\ref{app:symplecticPoisson}), and it is in this sense that we refer to
(\ref{PB}) as
a {symplectic realisation} of the magnetic monopole algebra~(\ref{a1}).

An important example of a symplectic realisation of the 
magnetic monopole algebra corresponds
to a constant and uniform magnetic charge distribution with density
$\vec\nabla\cdot\vec B=\rho$. In this paper we will study in detail
two particular examples of corresponding magnetic fields. A magnetic field with spherical symmetry is given
by
\bea\label{eq:Brot}
\vec B_{\rm spher}(\vec x\,)=\mbox{$\frac \rho3$}\, \vec x\ ,
\eea
with the symplectic algebra 
\bea
\{x^i,\bar\pi_j\}&=&\{\tilde x^i,\bar\pi_j\}\ \ =\ \ \{\tilde
x^i,\tilde\pi_j\}\ \ =\ \ \delta^i{}_j \ , \nn\\[4pt]
\{\bar\pi_i,\bar\pi_j\}&=&\mbox{$\frac{ e\,\rho}{3}$}\, \, \varepsilon_{ijk} \, \big(x^k-\tilde x^k\big)\ , \nn\\[4pt]	
\{\bar\pi_i,\tilde\pi_j\}&=&\{\tilde\pi_i,\bar\pi_j\} \ \ = \ \
\mbox{$\frac {e\,\rho}{6}$}\, \varepsilon_{ijk}\, x^k \ .\label{PB1}
\eea
As discussed in~\cite{Lust:2017bgx}, the uniform
magnetic charge density $\rho$ can be interpreted as a smearing of
infinitely many densely distributed Dirac monopoles of charge
$g=\rho$ and magnetic fields $\vec B_{\rm D}(\vec x-\vec
y\,)$, and in this setting there is a formal non-local magnetic vector
potential for \eqref{eq:Brot}. The gauge field on the extended
configuration space from \eqref{a4} in this case,
\bea\label{eq:Arot}
\vec A_{\rm spher}(x^I)=-\mbox{$\frac\rho6$}\,
\vec{\tilde x}\times\vec x \qquad \mbox{and} \qquad \vec{\tilde
  A}_{\rm spher}(x^I)=\vec0 \ ,
\eea
is a vector potential for \eqref{eq:Brot} in the sense of \eqref{eq:vecpotextended}. This gives a precise meaning to the absence of a local
definition of the vector potential and to the formal non-local smeared expression
derived in~\cite{Lust:2017bgx}: A definition of fields of
smooth magnetic charge distributions akin to classical Maxwell theory
necessitates the symplectic realisation of the 
magnetic monopole algebra. It is this feature that we shall exploit in the
following which makes the classical and quantum dynamics tractable.

At the opposite extreme, we can break spherical symmetry by choosing an
axial magnetic field which is oriented in the direction of a fixed
vector in $\real^3$; this is the analogue of a static uniform magnetic field
in the source-free case. We can choose coordinates in which this direction
is along the $z$-axis and take the linear magnetic field
\bea\label{eq:Baxial}
\vec B_{\rm axial}(\vec x\,) = (0,0,\rho\,z) \ ,
\eea
where $\vec x=(x,y,z)$. This leads to the symplectic algebra
\bea
\{x^i,\bar\pi_j\}&=&\{\tilde x^i,\bar\pi_j\}\ \ =\ \ \{\tilde
x^i,\tilde\pi_j\}\ \ =\ \ \delta^i{}_j \ , \nn\\[4pt]
\{\bar\pi_x,\bar\pi_y\}&=& e\,\rho\, z\ ,  \qquad
\{\bar\pi_x,\bar\pi_z\} \ \ = \ \ \mbox{$\frac{e\,\rho}2$}\, \tilde y
\qquad \mbox{and} \qquad \{\bar\pi_y,\bar\pi_z\} \ \ = \ \
-\mbox{$\frac{e\,\rho}2$}\, \tilde x \ ,\nn\\[4pt]	
\{\bar\pi_x,\tilde\pi_y\}&=&\{\tilde\pi_x,\bar\pi_y\} \ \ = \ \
\mbox{$\frac {e\,\rho}{2}$}\, z \ .\label{PB2}
\eea
The corresponding vector potential on the extended configuration space is
given by
\bea
\vec A_{\rm axial}(x^I)=\mbox{$\frac {\rho}{2}$}\, (-\tilde
y\, z,\tilde x\, z, 0) \qquad \mbox{and} \qquad \vec{\tilde A}_{\rm
  axial}(x^I)=\vec0\ .\label{eq:Aaxial}
\eea 

Both magnetic fields \eqref{eq:Brot} and \eqref{eq:Baxial} correspond
to a uniform magnetic charge distribution of density $\rho$, as
$\vec\nabla\cdot\vec B_{\rm spher}=\vec\nabla\cdot\vec B_{\rm
  axial}=\rho$, and their difference is a divergenceless magnetic
field
\bea
\vec B_{\rm axial}(\vec x\,)-\vec B_{\rm spher}(\vec x\,) = 
\vec\nabla\times\mbox{$\frac23$}\, \vec A_{\rm axial}(\vec x,\vec x\,) \ ,
\eea
which therefore does not contribute to the distribution of magnetic
charge. However, it does contribute to the twisted Poisson structure
defining the magnetic monopole algebra \eqref{a1}. Specifying only the
equation $\vec\nabla\cdot\vec B=\rho$ does not uniquely define the
corresponding magnetic field $\vec B$ so that, in contrast to
the magnetic field \eqref{eq:BDirac} of a Dirac monopole, here we
cannot apply considerations of spherical symmetry and the fact that
the field should be directed from the origin to fix the form of $\vec
B$. 
From a geometric perspective, the Poisson algebra \eqref{eq:piIpijcurl} is invariant
under the usual one-form gauge transformations $A_I\mapsto
A_I+\partial_I\chi$ on $\R^6$, as in this case $A_I(x^I)$ defines a
gauge field associated to a trivial $U(1)$-bundle over the extended
configuration space. However, it is not invariant under the higher two-form
gauge transformations $\vec B\mapsto\vec B+\vec\nabla\times\vec
a$ of the magnetic field on $\real^3$, which preserve the
curvature $\vec\nabla\cdot\vec B=\rho$. The essential geometric feature is that the field of a constant and uniform
magnetic charge distribution defines a ``higher'' gauge field
associated to a $U(1)$-gerbe on $\real^3$, contrary to the field of a Dirac
monopole which defines a gauge field of a non-trivial $U(1)$-bundle over
$\real^3\setminus\{\vec 0\,\}$ of degree $g$; see~\cite{Bunkinprep,Lust:2017bgx,SzaboISQS} for further discussion of
this point. This absence of higher gauge symmetry has profound
physical consequences, as we shall see later on.

\newsection{Classical dynamics from symplectic realisation\label{sec:classdyn}}

The classical motion of a spinless point particle of electric charge $e$ and mass $m$
under the influence of both a background magnetic field $\vec B(\vec
x\,)$ and a background
electric field $\vec E(\vec x\,)$ in three dimensions is governed by the Lorentz force
\bea
\frac{\dd\vec{\bar\pi}}{\dd t}=\frac em\, \vec{\bar\pi}\times\vec
B+e\, \vec E\ . \label{lf}
\eea
In this paper we assume that the electromagnetic backreaction due to
acceleration of the charged particle is negligible, and treat the
magnetic and electric fields in \eqref{lf} as fixed prescribed backgrounds.
As before, in this equation we need not assume that the magnetic field
$\vec B$ is divergenceless, i.e.~we include fields created by magnetic
poles. If $\vec\nabla\times\vec E=\vec0$, the electric field can be
represented as a gradient field $\vec E=\vec\nabla\phi$ for a globally
defined smooth scalar potential $\phi(\vec x\,)$ on $\real^3$. In this
case, the Lorentz force law \eqref{lf}, together with the definition $\dot{\vec
  x}=\vec{\bar\pi}/m$, can be written as the Hamilton equations of
motion $\dot x^i=\{x^i,\mcH\}$, $\dot{\bar\pi}_i=\{\bar\pi_i,\mcH\}$ with
the magnetic monopole algebra \eqref{a1} and the Hamiltonian taken to be the sum
of the kinetic energy and the electrostatic potential energy: $\mcH(\vec
x,\vec{\bar\pi}\,)= \vec{\bar\pi}^{\,2}/2m+e\,\phi(\vec x\,)$. When
$\vec\nabla\times\vec E \neq\vec0$, for instance when magnetic currents
are present, this is no longer possible when $\vec\nabla\cdot\vec B\neq0$. Here we shall allow for more
general electric fields in our formalism, i.e. we also allow for
motion in the field of 
general smooth distributions of electric poles. This is natural from
the present perspective, but we shall argue for it more precisely later on through
considerations of electromagnetic duality. The purpose of this
section is to demonstrate that the Lorentz
force \eqref{lf} follows in these generic situations from our
symplectic realisation of the magnetic monopole algebra by choosing an
appropriate Hamiltonian function $ \mcH$ on the extended phase space $\CCP$.

For later use, we shall momentarily leave the explicit form of the
vector potential $A_I(x^I)$ in \eqref{a3}
unspecified. Since the Poisson brackets \eqref{eq:piIpijcurl}
do not depend on momentum, while the Lorentz force is
linear in the kinematical momenta, the corresponding Hamiltonian is quadratic
in the momenta and of the general form
\bea 
 \mcH(x^I,\pi_I) =\frac 12\, \big(\begin{matrix} \vec\pi & \vec{\tilde\pi}
  \ \,
\end{matrix}\big) \cdot \bigg( \begin{matrix} a & b \\ b & c
\end{matrix}\bigg)\bigg( \begin{matrix} \, \vec\pi \ \, \\ \,
  \vec{\tilde\pi} \ \,
\end{matrix}\bigg) + V(x^I)\ ,\label{ham}
\eea
where $a$, $b$ and $c$ are real coefficients, and $ V(x^I)$ is a
smooth potential energy function on the extended configuration space. The
corresponding Hamilton equations of motion $\dot x^I=\{x^I, \mcH\}$,
$\dot\pi_I=\{\pi_I, \mcH\}$ read
\bea
\dot x^i&=&  a\, \pi^i+b\, {\tilde \pi}^i \ , \nn\\[4pt]
\dot {\tilde x}^i&=&b\, \pi^i+c\, {\tilde \pi}^i \ , \nn \\[4pt]
\dot \pi_i&=&\{\pi_i,\pi_j\}\, (a\, \pi^j+b\,
\tilde\pi^j)+\{\pi_i,\tilde\pi_j\}\, (b\, \pi^j+c\, {\tilde\pi}^j)-\p_iV \ , \nn\\[4pt]
\dot{\tilde \pi}_i&=&\{\tilde \pi_i,\pi_j\}\, (a\, \pi^j+b\,
{\tilde\pi}^j)+\{\tilde \pi_i,\tilde\pi_j\}\, (b\, \pi^j+c\,
{\tilde\pi}^j)-\tilde\p_iV \ , \label{heom}
\eea
from which one obtains the coupled system of second order differential equations for the
extended configuration space coordinates $x^i$ and $\tilde x^i$ given by
\bea
\ddot x^i &=& \{a\, \pi^i+b\, {\tilde\pi}^i,\pi_j\}\,\dot x^j +\{a\,
\pi^i+b\, {\tilde\pi}^i,\tilde\pi_j\}\,\dot {\tilde x}^j-a\, \p^iV- b\,
{\tilde\p}^iV \ , \label{eom1} \\[4pt]
\ddot {\tilde x}^i &=& \{b\, \pi^i+c\, {\tilde\pi}^i,\pi_j\}\,\dot x^j
+\{b\, \pi^i+c\, {\tilde\pi}^i,\tilde\pi_j\}\,\dot {\tilde x}^j- b\,
\p^iV+c\, {\tilde\p}^iV \ .\label{eom2}
\eea

If $a=0$, $b=2/m$, $\left\{\tilde\pi_i,\tilde\pi_j\right\}=0$ and
$\left\{\tilde\pi_i,\pi_j\right\}=(e/2)\, \varepsilon_{ijk}\, B^k(\vec
x\,)$,
which is exactly the case for the choice of vector potential
$A_I(x^I)$ in the form (\ref{a4}), then the terms proportional to the kinematical
momentum in the right-hand side of (\ref{eom1}) reproduce the correct
contribution $(e/m)\,\vec{\bar\pi}\times\vec B$ to the Lorentz force
from the magnetic field $\vec B(\vec x\,)$. The scalar potential
$ V(x^I)$ is then defined by the gradient field equation
\bea
-b\, \vec{\tilde\nabla} V=\mbox{$\frac em$} \, \vec E\ ,
\eea
which leads to
\bea\label{eq:scalarpotgen}
 V(x^I)=-\mbox{$\frac e2$}\, \big(\vec{\tilde x}\cdot\vec E(\vec x\,)+
\nu(\vec x\,)\big)\ ,
\eea
where $\nu(\vec x\,)$ is an arbitrary smooth function on
$\real^3$. An analogous ambiguity also appears in the definition of
the vector potential from \eqref{a4}: If we redefine $\vec A$ by
adding an arbitrary smooth vector field $\vec\alpha(\vec
x\,)=\vec{\tilde\nabla}\big(\vec{\tilde x}\cdot\vec\alpha(\vec x\,)\big)$ to get
\bea
\vec A(x^I)=\vec\alpha(\vec x\,)-\mbox{$\frac12$} \,\vec{\tilde
  x}\times\vec B(\vec x\,)\ ,\label{a4a}
\eea
then one still generates the magnetic field \eqref{eq:vecpotextended}
and the Poisson bracket
$\left\{\tilde\pi_i,\pi_j\right\}$ is unchanged, and hence so is the
corresponding equation of motion (\ref{eom1}). This 
ambiguity in the definition of the scalar and vector potentials will
be fixed below when we consider conditions for a consistent elimination of the auxiliary
degrees of freedom $(\vec{\tilde x},\vec{\tilde\pi}\,)$.

Setting $c=0$, the equations of motion thus become
\bea
\ddot x_i &=&\mbox{$\frac em$}\, \varepsilon_{ijk}\, \dot x^j\,
B^k(\vec x\,)+\mbox{$\frac em$}\,E_i(\vec x\,)\ , \label{eom3} \\[4pt]
\ddot {\tilde x}_i &=&\mbox{$\frac
  em$}\,\big[\p_i\alpha_j(\vec x\,)-\p_j\alpha_i(\vec x\,)+\big(\varepsilon_{ijk}\,
\p_lB^k(\vec x\,)-\varepsilon_{ijl}\, \p_kB^k(\vec x\,)\big)\, \tilde x^l\big]\,\dot x^j \nn\\
&&+\, \mbox{$\frac em$}\, \varepsilon_{ijk}\, \dot {\tilde x}^j\,
B^k(\vec x\,)+\mbox{$\frac em$}\,\big(\tilde x^j\, \p_iE_j(\vec
x\,)+\p_i\nu(\vec x\,)\big)\ .\label{eom4}
\eea
The corresponding Hamiltonian is
\bea
 \mcH(x^I,\pi_I) =\mbox{$\frac{1}{m}$}\, \pi_I\,\eta^{IJ}\,\pi_J
-\mbox{$\frac e2$}\, \big(\vec{\tilde x}\cdot \vec E(\vec x\,)+
\nu(\vec x\,)\big) \ , \label{Ham}
\eea
with the $O(3,3)$ metric
\bea
\eta^{IJ}=
\bigg( \begin{matrix}
0&\delta^i{}_j\\
\delta_i{}^j& 0
\end{matrix} \bigg) \ .
\eea
The Hamiltonian \eqref{Ham} is invariant under rotations in the dynamical symmetry group $O(3,3)\times O(3,3)$ of the extended phase space coordinates $(x^I,\pi_I)\in\real^6\times\real^6$ in the symplectic realisation.
The price to pay for the inclusion of generic magnetic fields $\vec
B(\vec x\,)$ and electric fields $\vec E(\vec x\,)$ here is the
presence of the auxiliary variables $\tilde x^i$, whose dynamics are
governed by the equation (\ref{eom4}). In the following we
will elucidate the physical meaning of the additional degrees of
freedom described by the coordinates $\tilde x^i$.

\newsection{Hamiltonian reduction\label{sec:Hamred}}

In this section we will view the original phase space $\CP$ as a
Hamiltonian reduction of the extended phase space $\CCP$ and analyse whether it is possible to consistently eliminate the
auxiliary variables from $\CCP$, in the sense of preserving the Lorentz force
equation \eqref{lf} for the observable coordinates $x^i$. 
We shall
answer this question in the negative: 
Upon imposing Hamiltonian constraints that get rid of the additional
degrees of freedom, our
model based on the symplectic realisation of the magnetic monopole
algebra cannot lead to a Lorentz force law describing the interaction with
magnetic charges and currents. 
More precisely, we show that introducing suitable
constraints recovers the standard model for the motion of electric
charge in a source-free magnetic field. However, these Hamiltonian
constraints annihilate the contribution to the Lorentz force from the magnetic
sources. In particular, for the spherically symmetric magnetic field \eqref{eq:Brot} they result in either free
particle motion, or in the absence of propagating degrees of freedom
altogether, i.e. the constrained mechanics is ``topological''.
This
exemplifies the role and necessity of the auxiliary
coordinates for a consistent Hamiltonian description of the interaction of the
electric charge with the background electromagnetic distributions.

In fact, it is rather elementary to see that the restriction of the dynamical
system with the Hamiltonian \eqref{Ham} to the constraint surface
$\CCC_0$ eliminates all propagating degrees of freedom. Conservation of the
primary constraint $\vec\phi=\vec{\tilde x}\approx\vec0$ gives
$\dot\phi^i=\{\phi^i,\mcH\}=\frac2m\, \pi^i\approx0$ and so results in the
secondary constraint $\vec\psi=\vec\pi\approx\vec0$. On the constraint
surface $\CCC_0$ all constraints $\Phi^I=(\phi^i,\psi_i)$ have
vanishing Poisson brackets among each other and are thus of first
class: $\{\Phi^I,\Phi^J\}\approx0$. But six first class constraints in
a $12$-dimensional phase space kills all dynamics and there are no
propagating degrees of freedom.

At this stage, however, we may ask whether, starting from the
symplectic realisation of the magnetic monopole algebra, there
is some more general constraint surface $\vec\phi(\vec x,\vec{\tilde
  x}\,)\approx\vec0$ and Hamiltonian \eqref{ham} such that the reduced
Hamiltonian dynamics reproduces the Lorentz force
\eqref{lf}. Let us
start again with a generic form for the vector potential
$A_I(x^I)$. The only way to obtain the Lorentz force
(\ref{lf}) from the system of differential equations (\ref{eom1}) and
\eqref{eom2} is via a linear primary constraint of the generic form
\bea
\vec\phi=\vec{\tilde x}-\zeta\, \vec x \approx\vec0 \ ,\label{c8}
\eea  
for some non-zero real parameter $\zeta$. We denote the corresponding constraint
surface in $\CCP$ by $\CCC_{\zeta}$ and global section of the
projection ${\sf p}:\CCP\to\CP$ by ${\sf s}_{\zeta}:\CP\to
\CCP$ with $
{\sf s}_{\zeta}^*\,\Omega=\Omega|_{\CCC_{\zeta}}$. 
Conservation of the primary constraint
(\ref{c8}) implies the strong relations
\bea
\frac{\dd\phi^i}{\dd t} = \{\phi^i,\mcH\}= (\zeta\,a-b)\,
\pi^i+(\zeta\,b - c)\,\tilde\pi^i\approx0\ ,\label{a9}
\eea
and the analysis of the constrained Hamiltonian dynamics now depends
on which region we consider of the four-dimensional space of parameters
$(a,b,c,\zeta)$.

Suppose first that the parameters satisfy
$\zeta=\frac ba=\frac cb$. Then \eqref{a9}
is identically zero, so \eqref{c8} are first class constraints
in this case. The Hamiltonian \eqref{ham} becomes
\bea
\mcH(x^I,\pi_I) =\mbox{$\frac
  a2$}\,\big(\vec\pi+\zeta \,\vec{\tilde\pi}\,\big)^2 +
V(x^I) \ .
\eea
To obtain the constrained Hamilton equations of motion we introduce
the total Hamiltonian $\mcH_{\rm tot}=\mcH+\vec u\cdot\vec\phi$, where $u_i$
are Lagrange multipliers, and write $\dot f\approx
\sfs_{\zeta}^*\{f,\mcH_{\rm tot}\}$ on the constraint surface
$\CCC_{\zeta}$ for any function $f$ on the extended phase space
$\CCP$. Using \eqref{heom} the equations of motion for the phase space
coordinates now read
\bea
\dot x^i &\approx& a\,\big(\pi^i+\zeta \,
\tilde\pi^i\big) \ , \nn\\[4pt]
\dot\pi_i &\approx&
a\,\{\pi_i,\pi_j\}_\zeta\,\big(\pi^j+\zeta \,
\tilde\pi^j\big) + a\,\zeta \,
\{\pi_i,\tilde\pi_j\}_\zeta\big(\pi^j+\zeta \,
\tilde\pi^j\big) - \partial_iV_\zeta-\zeta\,u_i \ , \nn\\[4pt]
\dot{\tilde x}^i &\approx& a\, \zeta \,
\big(\pi^i+\zeta \, \tilde\pi^i\big) \ , \nn\\[4pt]
\dot{\tilde\pi}_i &\approx& a\,
\{\tilde\pi_i,\pi_j\}_\zeta\, \big(\pi^j+ \zeta \,
\tilde\pi^j\big) + a\, \zeta \,
\{\tilde\pi_i,\tilde\pi_j\}_\zeta\,
\big(\pi^j+ \zeta \, \tilde\pi^j\big) +\mbox{$\frac1\zeta$}\, \partial_iV_\zeta+ u_i \ ,
\label{c1a}\eea
where $\{\pi_I,\pi_J\}_\zeta:=\sfs_\zeta^*\{\pi_I,\pi_J\}$ and
$V_\zeta(\vec x\,):= V(\vec x,\zeta\,\vec x\,)$.
We now observe that the combination of covariant momenta
$\Pi_i:=\pi_i+\zeta\, \tilde\pi_i$ is conserved,
$\dot{\Pi}_i=0$, with $\dot{ x}^i=a\,\Pi^i$ and
$\dot{\tilde{ x}}^i=a\, \zeta\,\Pi^i$, which implies
$\ddot{ x}^i=\ddot{\tilde{ x}}^i=0$. Thus instead of the
Lorentz force due to the electromagnetic field, we obtain free
particle motion for the configuration space degrees of freedom $\vec
x$ and $\vec{\tilde x}$. That is, this region of parameter space is
not suitable for our aims.

Henceforth we therefore assume $\zeta\neq\frac cb$. Then (\ref{a9}) can be represented as a secondary constraint
\bea
\vec\psi=\vec{\tilde\pi}-\gamma\, \vec{\pi}\approx\vec0 \ , \label{c9}
\eea
where
\bea\label{eq:gammadef}
\gamma=-\frac{\zeta\,a- b}{\zeta \, b- c} \ . \label{c10}
\eea
Let us now write the total
Hamiltonian $\mcH_{\rm tot}=\mcH+\vec u\cdot\vec\phi+\vec v\cdot\vec\psi$,
and solve for the Lagrange
multipliers $u_i$ and $v^i$. The Poisson brackets of the constraints are
$\{\psi_i,\phi^j\}=-(1+\zeta\,\gamma) \,\delta_i{}^j$, 
and we suppose that $\zeta\,\gamma \neq -1$, or equivalently $a\, \zeta^2-2\,b\,\zeta+c
\neq 0$, for otherwise the constraints are of first class implying the
absence of propagating degrees of freedom.
Then from the strong equality
$\{\phi^i,\mcH_{\rm tot}\}\approx0$ one finds $\vec v=\vec0$. Conservation
of the constraint \eqref{c9} implies the
strong relation $\{\psi_i,\mcH_{\rm tot}\}\approx0$, from which one thereby
obtains
\bea
u_i=\mbox{$\frac{1}{1+\zeta\,\gamma}$} \, \big( (a+\gamma\,b)\, 
\{\tilde\pi_i-\gamma\,\pi_i , 
\pi_j+\zeta \, \tilde\pi_j\}_\zeta \, \pi^j
 +(\mbox{$\frac1\zeta$}-\gamma)
\, \partial_iV_\zeta\big)
\ ,
\eea
where we used the identity $b+\gamma\,c=\zeta\, (a+\gamma\,b)$
that follows from the definition \eqref{eq:gammadef}.
The resulting equations of motion are given by
\bea 
 \dot x^i \ \ \approx \ \ \sfs_{\zeta}^*\{x^i,\mcH_{\rm tot}\} & = &
 (a + \gamma\, b)\, \pi^i \ , \nn\\[4pt]
 \dot \pi_i \ \ \approx \ \ \sfs_{\zeta}^*\{\pi_i,\mcH_{\rm tot}\}
 & = & \mbox{$\frac{1}{1+\zeta\, \gamma}$}\, \big( (a+\gamma\,b)\, 
 \{ \pi_i+\zeta\,\tilde\pi_i,
 \pi_j+\zeta \, \tilde\pi_j\}_\zeta \, \pi^j -
 (2-\zeta\,\gamma)\, \partial_iV_\zeta\big)\ , \label{c14}
\eea
which lead to the second order differential equations
\bea
\ddot x^i = \mbox{$\frac{a+\gamma\,b}{1+\zeta\, \gamma}$} \, \big(
 \{\pi^i+\zeta\,\tilde\pi^i,
 \pi_j+\zeta\, \tilde\pi_j\}_\zeta \, \dot x^j -
 (2-\zeta\,\gamma)\, \partial^iV_\zeta\big)
 \ .\label{eom5}
\eea
This is exactly the Lorentz force corresponding to the effective magnetic field
\bea
B_{\rm eff}^i=\mbox{$\frac me\, \frac{a+\gamma\,b}{1+\zeta\, \gamma}$}
\
  \varepsilon^{ijk} \, \big(
 \{ \pi_j+\zeta\,\tilde\pi_j,
 \pi_k+\zeta\, \tilde\pi_k\}_\zeta \big) \ ,\label{a13}
\eea
and the effective electric field
\bea
\vec E_{\rm eff} = -\mbox{$\frac me\, \frac{(a+\gamma\,b)\,
    (2-\zeta\,\gamma)}{1+\zeta\, \gamma}$} \ \vec\nabla V_\zeta \ .\label{Eeff}
\eea

By explicitly calculating the pullback of the Poisson brackets 
\eqref{eq:piIpijcurl} (as a
two-form) to the constraint surface $\CCC_\zeta$, the 
magnetic field \eqref{a13} can be written in the form
\bea\label{aeff}
\vec B_{\rm eff} = \mbox{$\frac{m\,(\zeta+1)\,
    (a+\gamma\,b)}{1+\zeta\,\gamma}$} \ \vec\nabla\times\big(\vec
A_\zeta+\zeta^2\,\vec{\tilde A}_\zeta\big) \ ,
\eea
where $\vec A_\zeta(\vec x\,):=\vec A(\vec x,\zeta\,\vec x\,)$. Since
the effective magnetic field is derived from an effective vector
potential, it satisfies $\vec\nabla\cdot\vec B_{\rm eff}=0$ and so
cannot be sourced by monopoles. Writing $\vec B_{\rm mag}=\vec B-\vec
B_{\rm eff}$, we can decompose the original magnetic
field $\vec B$ as
\bea\label{dec}
\vec B = \vec B_{\rm mag}+ \mbox{$\frac{m\,(\zeta+1)\,
    (a+\gamma\,b)}{1+\zeta\,\gamma}$} \ \vec\nabla\times\big(\vec
A_\zeta+\zeta^2\,\vec{\tilde A}_\zeta\big) \ ,
\eea
where the magnetic field $\vec B_{\rm mag}$ with $\vec\nabla\cdot\vec
B_{\rm mag}=\vec\nabla\cdot\vec B$ accounts for the contributions from
magnetic charge distributions, while \eqref{aeff} is the magnetic
field created by electric currents and time-varying electric
fields. For the Hamiltonian \eqref{Ham} and the specific choice of
vector potential $A_I(x^I)$ from \eqref{a4a}, one has
\bea\label{Beff}
\vec B_{\rm eff}(\vec x\,) = \mbox{$\frac{\zeta+1}{2\,\zeta}$} \
\vec\nabla\times\big(\vec\alpha(\vec x\,) - \mbox{$\frac\zeta2$}\, \vec x\times\vec B(\vec x\,)\big) \ .
\eea
Setting $\vec\alpha=\vec0$, for both spherically symmetric magnetic fields \eqref{eq:BDirac} of a Dirac monopole
and \eqref{eq:Brot} of a uniform magnetic charge
distribution, the corresponding effective magnetic field \eqref{Beff}
vanishes and $\vec B_{\rm mag}=\vec B$; in these cases, the
constrained dynamics describes free particle motion in the absence of
a force due to the potential $V_\zeta$. On the other hand, for the
axial magnetic field \eqref{eq:Baxial} with $\zeta=1$ we find $\vec
B_{\rm eff}=\frac32\, (\vec
B_{\rm axial}-\vec B_{\rm spher})$ and $\vec B_{\rm
  mag}=\frac12\,(3\,\vec B_{\rm spher}-\vec B_{\rm axial})$.

Likewise, since the effective electric field \eqref{Eeff} is a
gradient field of an effective electrostatic charge distribution, it satisfies
$\vec\nabla\times \vec E_{\rm
  eff}=\vec0$ and hence cannot be
sourced by magnetic currents. Writing $\vec E_{\rm mag}=\vec E-\vec
E_{\rm eff}$, we can decompose the original electric field $\vec E$ as
\bea
\vec E = \vec E_{\rm mag} -\mbox{$\frac me\, \frac{(a+\gamma\,b)\,
    (2-\zeta\,\gamma)}{1+\zeta\, \gamma}$} \ \vec\nabla V_\zeta \ ,
\eea
where the electric field $\vec E_{\rm mag}$ with $\vec\nabla\times\vec
E_{\rm mag} = \vec\nabla\times\vec E$ accounts for the contributions
from magnetic currents and time-varying magnetic fields, while
\eqref{Eeff} is the electric field sourced by electric charge
distributions. For the Hamiltonian \eqref{Ham} one has
\bea\label{Eeff1}
\vec E_{\rm eff}(\vec x\,) = \mbox{$\frac1{2\zeta}$} \
\vec\nabla\big(\zeta\, \vec
x\cdot\vec E(\vec x\,) + \nu(\vec x\,)\big) \ .
\eea

We can now ask for which original magnetic fields $\vec B$ and
electric fields $\vec E$ do the constrained equations of motion
(\ref{eom5}) coincide with the original equations from (\ref{eom3}), or equivalently when do the effective magnetic and electric fields from
\eqref{Beff} and \eqref{Eeff1} (with $\zeta=1$) coincide with the original fields: $\vec
B_{\rm eff}=\vec B$ and $\vec E_{\rm eff}=\vec E$. For the magnetic
field, one firstly requires $\vec\nabla\cdot\vec B=0$, so there exists
a magnetic vector potential $\vec a(\vec x\,)$ with $\vec
B=\vec\nabla\times\vec a$. From the identity
\bea
\vec\nabla\times(\vec
x\times\vec B\,)=(\vec\nabla\cdot\vec B\,)\, \vec x-(\vec
x\cdot\vec\nabla)\vec B-2\,\vec B \ ,
\eea
it follows that this requires
$\vec\nabla\times\vec\alpha = -\frac12\,(\vec x\cdot\vec\nabla)\vec
B$. Since the vector field $\vec\alpha$ in the definition of the vector
potential $\vec A$ from (\ref{a4a}) is arbitrary, we can take
\bea\label{a15}
\vec\alpha(\vec x\,)=\vec a(\vec x\,)+\mbox{$\frac12$}\, \vec
x\times\vec B(\vec x\,) \ .
\eea
It is useful to retain the contribution (\ref{a15}) to the vector
potential (\ref{a4a}) even in the more general case whereby $\vec\nabla\cdot \vec B\neq0$,
and using the decomposition (\ref{dec}) we therefore write
\bea
\vec A(x^I)=\vec a(\vec x\,)-\mbox{$\frac12$}\, \vec x\times\vec
B_{\rm mag}(\vec x\,)-\mbox{$\frac12$}\, \big(\vec{\tilde x}-\vec x\,
\big)\times\vec B(\vec x\,)\ ,\label{vecp}
\eea
where here $\vec\nabla\times\vec a=\vec B_{\rm eff}$. For the
spherically symmetric field \eqref{eq:Brot} the vector potential
\eqref{vecp} coincides with \eqref{eq:Arot}, while for the axial field
\eqref{eq:Baxial} it modifies \eqref{eq:Aaxial} to
\bea
\vec A_{\rm axial}(x^I)=\mbox{$\frac{\rho}{4}$}\, z\, (y-2\tilde
y,2\tilde x-x, 0) \qquad \mbox{and} \qquad \vec{\tilde A}_{\rm axial}(x^I)=\vec0\
. \label{eq:Aaxialmod}
\eea 

In a completely
analogous way, we use the arbitrariness of the function $\nu(\vec
x\,)$ appearing in the definition the scalar potential $V(x^I)$ from
\eqref{eq:scalarpotgen} to
ensure that the
effective electric field $\vec E_{\rm eff}$ from (\ref{Eeff}) coincides with the original
electric field $\vec E$ when
$\vec\nabla\times\vec E=\vec0$. In the general case we thus find
\bea
V(x^I)=e\, \phi(\vec x\,)-\mbox{$\frac e2$} \, \vec
x\cdot\vec E_{\rm mag}(\vec x\,)-\mbox{$\frac e2$}\, \big(\vec{\tilde
  x}-\vec x\,\big)\cdot\vec E(\vec x\,) \ ,\label{scalp}
\eea
where $\vec\nabla\phi=\vec E_{\rm eff}$.

We have thereby established that, independently of the choice of the
vector potential $A_I(x^I)$ and Hamiltonian $\mcH(x^I,\pi_I)$, imposition of the Hamiltonian constraints
\eqref{c8} leads to the Lorentz force with the source-free
magnetic and electric fields $\vec B_{\rm eff}$ and $\vec E_{\rm
  eff}$. Conversely, if the original magnetic and electric fields are
source-free then the choices (\ref{vecp}) for the vector potential
$A_I(x^I)$ and \eqref{scalp} for the scalar potential $V(x^I)$ ensure
that the constrained equations of motion (\ref{eom5}) coincide with
the original Lorentz force (\ref{eom3}). Thus \emph{only} in
this source-free case can we eliminate the auxiliary variables via Hamiltonian
reduction, and thus recover the standard model for the dynamics of
electric charges in magnetic and electric fields.
From a geometric perspective we have shown that, by fixing the
equations of motion \eqref{lf} as the fundamental entities, there is
no polarisation on the extended symplectic algebra which is compatible
with both the Lorentz force law and nonassociativity of the magnetic
monopole algebra \eqref{a1}: No polarisation can lead to
a nonassociative algebra, and only associative algebras are possible
upon Hamiltonian reduction.

\newsection{Dyonic motion\label{sec:Sduality}}

Thus far our considerations have treated magnetic and electric fields
on almost equal footing, and it is natural to extend our formalism in
a way which incorporates the electromagnetic fields symmetrically. In
fact, one of the arguments supporting the existence of magnetic
monopoles is the desire to extend the electromagnetic duality of
vacuum Maxwell theory to cases with sources. Recall that the
electromagnetic duality transformation is the map of order four
acting on electric and magnetic fields as
\bea\label{emd}
(\vec E,\vec B\,)\longmapsto (\vec B,-\vec E\,) \ .
\eea
This transformation generates a cyclic subgroup $\Z_4\subset SO(2)$ of
the global
symmetry group of Maxwell theory consisting of electromagnetic duality rotations
\bea\label{eq:Sduality}
\bigg( \begin{matrix}
\, \vec E \ \, \\ \, \vec B \ \,
\end{matrix} \bigg)
\longmapsto
\bigg( \begin{matrix}
\cos\theta & \sin\theta \\ -\sin\theta & \cos\theta
\end{matrix} \bigg)
\bigg( \begin{matrix}
\, \vec E \, \ \\ \, \vec B \ \,
\end{matrix} \bigg) \ ,
\eea
with $\theta\in[0,2\pi)$, for which \eqref{emd} is the
$\theta=\frac\pi2$ member of this family of continuous symmetries.

If a point particle of mass $m$ is a dyon with electric and magnetic charges $q_e$ and $q_m$, respectively, the corresponding Lorentz force law becomes
\bea
\frac{\dd\vec{\bar\pi}}{\dd t}=\frac1m\, \vec{\bar\pi}
\times\big(q_e\, \vec B-q_m\, \vec E\,\big)+q_e\, \vec E+q_m\, \vec B\
. \label{lfem}
\eea
This equation is invariant under the electromagnetic duality rotations
(\ref{eq:Sduality}) if the charges of the dyon also
transform correspondingly as
\bea\label{eq:chargerot}
\bigg( \begin{matrix}
q_e \\ \, q_m \,
\end{matrix} \bigg)
\longmapsto
\bigg( \begin{matrix}
\cos\theta & \sin\theta \\ -\sin\theta & \cos\theta
\end{matrix} \bigg)
\bigg( \begin{matrix}
q_e \\ \, q_m \,
\end{matrix} \bigg) \ .
\eea
A Hamiltonian formalism for the equations of motion \eqref{lfem} can
be developed along the lines of Sections~\ref{sec:classdyn} and~\ref{sec:Hamred}, by simply substituting everywhere the original
electric and magnetic fields with the corresponding combinations
\bea
e\, \vec E \longrightarrow q_e \, \vec E+q_m\, \vec B \qquad
\mbox{and} \qquad e\, \vec B \longrightarrow q_e\, \vec B-q_m\, \vec E \ .
\eea
Our symplectic
realisation circumvents the usual problems of electromagnetic duality associated with relating
dual vector potentials locally with the original ones. Let us look at
two explicit examples in detail. 

Consider first the interaction of a pair of dyons in three dimensions.
We consider the field \eqref{eq:BDirac} of a Dirac monopole, for which
the effective magnetic field vanishes and $\vec B_{\rm mag}=\vec B_{\rm D}$, and we also introduce the electric field
\bea
\vec E_{\rm C}(\vec x\,) = e\ \frac{\vec x}{|\vec x\,|^3}
\eea
corresponding to the Coulomb force exerted by a point charge. Then the prescription of Section~\ref{sec:Hamred} yields the corresponding vector and scalar potentials
\bea
\vec A_{\rm D}(x^I)=-\frac {q_e\,g-q_m\, e}{2\,|\vec x\,|^3}\ {\vec {\tilde x}}\times\vec x \qquad \mbox{and} \qquad V_{\rm C}(x^I)=-\frac{q_e\,e+q_m\,g}{2\,|\vec x\,|^3}\ \vec x\cdot\big(\vec x+\vec{\tilde x}\, \big) \ .
\eea
Note that, in contrast to \eqref{eq:aDirac}, the vector potential on
the extended configuration space has no Dirac string singularities and
is defined for all $(\vec x,\vec{\tilde
  x}\,)\in\big(\real^3\setminus\{\vec0\,\}\big)\times\real^3$, which
coincides with the domain of the magnetic field \eqref{eq:BDirac} in $\real^6$. 
The Hamiltonian describing the interaction of two dyons of charges $(q_m,q_e)$ and $(g,e)$ is then given by
\bea
\mcH_{\rm DC}(x^I,p_I)&=&\frac{2}{m}\
{\vec\pi}_{\rm D}\cdot{\vec{\tilde\pi}}_{\rm D}+V_{\rm C}(x^I)\nn\\[4pt] &=&\frac{2}{m}\ {\vec
  p}\cdot{\vec{\tilde p}}-\frac{1}{m\,|\vec x\,|^2} \
\big(\vec {\tilde x}\times\vec J \ \big)\cdot\vec{\tilde p}-
\frac{q_e\, e + q_m\, g}{2\,|\vec x\,|^3} \ \vec x\cdot\big(\vec x+\vec{\tilde x}\, \big)\ ,
\eea
where the vector
\bea
\vec J = -\big(q_e\,g- q_m\,e\big)\ \frac{\vec x}{|\vec x\,|}
\eea
is the angular momentum of the electromagnetic field produced by the
pair of dyons around the axis through the midpoint separating them. In
the quantum mechanics that we consider in
Section~\ref{sec:quantumdynamics}, the components of the total angular
momentum operator generate the rotation group $SU(2)$, and quantum
states thereby form
representations of $SU(2)$. Requiring that they generate a
finite-dimensional representation of $SU(2)$ leads to the quantisation
of angular momentum, giving
\bea\label{eq:Diracquantdyon}
q_e\,g- q_m\,e = 
\mbox{$\frac\hbar2$}\, n \qquad \mbox{with} \quad n\in\Z \ ,
\eea
which is just Dirac charge quantisation~\cite{Dirac1931,Dirac1948}. The quantisation condition \eqref{eq:Diracquantdyon} is preserved by electromagnetic duality rotations \eqref{eq:chargerot} of both sets of dyon charges.

Consider next the motion of a single dyon in the spherically symmetric fields of constant and uniform magnetic and electric charge distributions of densities $\rho_m$ and $\rho_e$, respectively. We assume there are no currents and set
\bea
\vec B_{\rm spher}=\mbox{$\frac{\rho_m}{3}$}\, \vec x \qquad \mbox{and} \qquad \vec E_{\rm spher}=\mbox{$\frac{\rho_e}{3}$}\, \vec x\ .
\eea
The Lorentz force law reads
\bea
\frac{\dd\vec{\bar\pi}}{\dd t}=\frac{q_e\, \rho_m-q_m\, \rho_e}{3m} \
\vec{\bar\pi}\times\vec x+\frac{q_e\, \rho_e+q_m\, \rho_m}{3} \ \vec x\ .\label{lfem1}
\eea
The duality rotation in this case is given by \eqref{eq:chargerot}
together with
\bea
\bigg( \begin{matrix}
\rho_e \\ \, \rho_m \,
\end{matrix} \bigg)
\longmapsto
\bigg( \begin{matrix}
\cos\theta & \sin\theta \\ -\sin\theta & \cos\theta
\end{matrix} \bigg)
\bigg( \begin{matrix}
\rho_e \\ \, \rho_m \,
\end{matrix} \bigg) \ .\label{emd1}
\eea
Then the equations of motion (\ref{lfem1}) are clearly invariant under
the transformations \eqref{eq:chargerot}
and (\ref{emd1}), as the rotation group $SO(2)$ preserves both the natural
symplectic form $\vec q\wedge\vec\rho$ and inner product $\vec
q\cdot\vec\rho$ on the two-dimensional vector space $\real^2$. Following the prescription of
Section~\ref{sec:Hamred}, we find the vector and scalar potentials
\bea
\vec A_{\rm spher}(x^I)=-\frac{{q_e\choose \, q_m \, }\wedge{\rho_e\choose \,
    \rho_m \, }}{6} \ \vec {\tilde
  x}\times\vec x \qquad \mbox{and} \qquad V_{\rm spher}(x^I)=-\frac{{q_e\choose \,
    q_m \, }\cdot{\rho_e\choose \, \rho_m \, }}{6} \ \vec{\tilde x}\cdot\vec x\ ,
\eea
and we write the corresponding Hamiltonian on the extended phase space
$\CCP$ as
\bea
\mcH_{\rm spher}(x^I,p_I)&=&\frac{2}{m}\
{\vec\pi}_{\rm spher}\cdot{\vec{\tilde\pi}}_{\rm spher}+V_{\rm spher}(x^I)\nn\\[4pt] &=&\frac{2}{m}\ {\vec
  p}\cdot{\vec{\tilde p}}+\frac{{q_e\choose \, q_m \, }\wedge{\rho_e\choose \,
    \rho_m \, } }{3m} \
\big(\vec {\tilde x}\times\vec x\, \big)\cdot\vec{\tilde p}-
\frac{{q_e\choose \, q_m \, }\cdot{\rho_e\choose \,
    \rho_m \, }}{6} \ \vec{\tilde x}\cdot\vec x\ .
\label{eq:Hspherdyon}\eea

\newsection{Integrability\label{sec:integrability}}

Let us now address the problem of integrating the Lorentz force
equation \eqref{lf}. The symplectic realisation of the magnetic monopole algebra may be
used to consistently formulate the time evolution of classical
observables; the Jacobi identity together with the Leibniz rule allows
for the implementation of the classical Liouville theorem to construct
integrals of motion in principle. We look at this in detail for two particular
electromagnetic backgrounds in turn.

\bigskip

\noindent
{\bf Spherically symmetric fields. \ } 
We consider first the Hamiltonian
\eqref{Ham} with the spherically symmetric magnetic field
\eqref{eq:Brot} and no electric background, $\vec
E=\vec0$, which is given by
\bea
\mcH_{\rm spher}(x^I,p_I)=\frac{2}{m} \, \vec\pi_{\rm spher}\cdot\vec{\tilde\pi} _{\rm spher} = \frac2m\ {\vec
  p}\cdot{\vec{\tilde p}} + \frac{e\,\rho}{3m} \
\big(\vec {\tilde x}\times\vec x\, \big)\cdot\vec{\tilde p} \ .
\eea
The solutions of the classical equations of motion
\eqref{eom3} and \eqref{eom4} are
the union of the integral curves of the vector field $\frac2m\, \big(
\vec{\tilde\pi}\,,\,\vec\pi\,,\,\frac{e\,\rho}{6}\,(
\vec\pi\times\vec x - 2\,\vec{\tilde\pi}\times\vec{\tilde x} \,
)\,,\,\frac{e\,\rho}{6}\, \vec{\tilde\pi}\times\vec
x\,\big)$ on the extended phase space $\CCP$ with the corresponding Hamiltonian flow equations
\bea \label{eq:characteristics}
\dot{\vec x}=\mbox{$\frac2m$}\, \vec{\tilde\pi}\ , \qquad
\dot{\vec{\tilde x}}=\mbox{$\frac2m$}\, \vec\pi \ , \qquad 
\dot{\vec\pi}=\mbox{$\frac{e\,\rho}{3m}$}\,
\big(\vec\pi\times\vec x - 2\,\vec{\tilde\pi}\times\vec{\tilde x}\,
\big) \qquad \mbox{and} \qquad \dot{\vec{\tilde\pi}}= \mbox{$\frac{e\,\rho}{3m}$}\,
\vec{\tilde\pi}\times\vec x \ .
\eea

We need to find from these flow equations a sextuple of integrals of
motion $(\CI_1,\dots,\CI_6)$, i.e.~$\dot \CI_I=\{\CI_I,\mcH\}=0$ for $I=1,\dots,6$. Several integrals of motion are
readily found: As usual, the Hamiltonian 
\bea
\CI_1=\mcH_{\rm spher} 
\eea
is trivially conserved, and so is the kinetic energy
\bea
\CI_2=\mbox{$\frac2{m}$} \, \vec{\tilde\pi}^{\,2}
\eea
which follows easily from \eqref{eq:characteristics} as $\dot
\CI_2=\frac4m\, \vec{\tilde\pi}\cdot\dot{\vec{\tilde\pi}}=0$. There is also the azimuthal angular momentum
\bea
\CI_3=-\vec L^{\,2} \ ,
\eea
which corresponds geometrically to the volume of the tetrahedron
$\triangle(\vec x,2\,\vec{\tilde\pi},\vec L\, )$ in
the extended phase space $\CCP$ 
formed by the position vector $\vec x$, the kinematical momentum
$2\,\vec{\tilde\pi}$, and the orbital angular momentum $\vec L=2\,\vec
x\times \vec{\tilde\pi}$ of the charged particle. The proof that $\CI_3=-4\,(\vec x\times \vec{\tilde\pi}\, )^2$ is a conserved quantity
easily follows from the triple scalar product identity
\bea
\CI_3= 2\,\vec x\cdot (\vec L\times\vec{\tilde\pi}\, ) = 4\,(\vec
x\cdot \vec{\tilde\pi}\, )^2 -4\,\vec x^{\,2}\, \vec{\tilde\pi}^{\,2}
\eea
together with \eqref{eq:characteristics}.

However, in addition to commuting with the Hamiltonian $\mcH$, these
three integrals of motion are in involution with each other and
therefore do not produce any new conserved quantities. We have not been
able to find another three integrals of motion that would enable the
integration of the Hamilton equations of motion on the extended phase
space $\CCP$. This problem is also considered directly in the original
phase space $\CP$ by~\cite{BL} where it is
suggested that, despite its spherical symmetry, the Lorentz force
equations \eqref{lf} in the magnetic field \eqref{eq:Brot} do not
appear to be integrable. This is in marked
contrast to the case of the magnetic field \eqref{eq:BDirac} of a
Dirac monopole, for which the Hamiltonian on extended phase space becomes
\bea\label{eq:HDiracext}
\mcH_{\rm D}(x^I,p_I)=\frac{2}{m}\
{\vec\pi}_{\rm D}\cdot{\vec{\tilde\pi}}_{\rm D}=\frac{2}{m}\ {\vec
  p}\cdot{\vec{\tilde p}}+\frac{e\, g}{m\,|\vec x\,|^3}\ \big(\vec {\tilde x}\times\vec x\, \big)\cdot\vec{\tilde p}\ .
\eea
In this case integrability is ensured by conservation of the
Poincar\'e vector
\bea
\vec K=\frac 2m\ \vec x\times\vec{\tilde p}-\frac{e\, g}{m}\ \frac{\vec x}{|\vec x\,|}\ ,
\eea
which is proportional to the sum of the orbital angular momentum $\vec
L$ with the angular momentum of the electromagnetic field due to the
electric charge and the Dirac monopole; one easily checks that the components of $\vec K$ 
commute with the extended Hamiltonian \eqref{eq:HDiracext} and also with the kinetic energy $\frac2m\, \vec{\tilde\pi}^{\,2}$. In
particular, the Dirac
charge quantisation condition \eqref{eq:Diracquantdyon} in this case simplifies to
\bea\label{eq:Diracquant}
e\, g = \mbox{$\frac\hbar2$}\, n \qquad \mbox{with} \quad n\in\Z \ .
\eea
The conservation of the Poincar\'e
vector ensures that the charged particle never reaches the location of
the monopole~\cite{BL}, as it precesses around the direction $\vec K$
with time-varying angular frequency and the motion is confined to the
surface of a cone whose apex is the location of the monopole.\footnote{This is of course a well-studied system, and many extensions and reductions have been considered previously, see e.g.~\cite{Plyushchay2000,Plyushchay2013}.}

It was shown in~\cite{BL}
that the motion of an electric charge in the magnetic field
\eqref{eq:Brot} can be effectively described as the dynamics in the
field of a single Dirac monopole with some frictional force: After a suitable time reparameterisation, the Lorentz force \eqref{lf} can be brought to the form
\bea
m\,\ddot{\vec x}+\lambda(t)\, \dot{\vec x} = e\, \dot{\vec x}\times\vec B_{\rm D} \ ,
\eea
where the time-dependent friction coefficient $\lambda(t)$ captures the uniform distribution of magnetic charge. In particular, the motion is no longer confined in any direction.
This
interpretation lends a physical explanation for the necessity of keeping
auxiliary degrees of freedom in order to reproduce the correct equations of
motion \eqref{lf} as we demonstrated in Section~\ref{sec:Hamred}: A
consistent Hamiltonian description of dissipative dynamics with
friction typically requires the introduction of additional degrees of freedom
describing the reservoir which is needed to absorb the dissipated energy, see Appendix~\ref{app:dho}. This analogy will be especially prominent
when we consider the quantisation of this system below.
For dissipative systems the auxiliary degrees of freedom are needed to
conserve the total energy. In the present case the energy is already conserved
in the ``physical'' sector, suggesting that there may be another
physical quantity which is not conserved in the physical subsystem but
only in the complete doubled system. It would be interesting to understand this further in order to better clarify the physical
meaning of the auxiliary coordinates in our case.

\bigskip

\noindent
{\bf Axial fields. \ } 
The situation is remarkably simpler in the case of the axial magnetic
field \eqref{eq:Baxial}. Let us first study the dynamics of the
physical coordinates. The Lorentz force in components from \eqref{eom3} reads 
\bea
\dot{\tilde\pi}_x=\omega \, z \, {\tilde\pi}_y\ ,\qquad
\dot {\tilde\pi}_y=-\omega \, z\, {\tilde\pi}_x\qquad \mbox{and} \qquad
\dot {\tilde\pi}_z=0\ ,\label{m1}
\eea
where $\omega=e\,\rho/m$ and we assume here that $e\,\rho>0$. From the third equation we discover another
integral of motion given by the kinematical
momentum in the direction of the magnetic field, and from it we get $z(t)=v_z\,
t+z_0$. With the appropriate choice of origin of coordinates, we may
set the initial position to $z_0=0$. $v_z$ is the constant velocity in
the $z$-direction, and we suppose that $v_z>0$. Note that the Lorentz
force in (\ref{m1}) is different from the force exerted by the time-dependent magnetic field $\vec B_t=(0,0,v_z\,\rho\, t)$ that would create an electric field $\vec E=v_z\,\rho\,(y,-x,0)$, which is absent from (\ref{m1}). We will incorporate an electric background below to properly simulate dyonic motion, but for the moment we focus on the solutions to the system \eqref{m1}.

\begin{figure}[htb]
\centering
  \includegraphics[width=3.5cm,height=6.5cm]{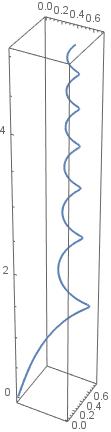}
\caption{\small Motion of an electric charge in a linear magnetic
  field along a fixed vector.}\label{fig:axialmotion}
\end{figure}

From the second equation we then find
${\tilde\pi}_x=-\dot {\tilde\pi}_y/\omega\, v_z\, t$, and so the first equation gives
\bea
\ddot {\tilde\pi}_y-\mbox{$\frac 1t$}\, \dot {\tilde\pi}_y+\omega^2\, v_z^2\, t^2 \, {\tilde\pi}_y=0\ .\label{m2}
\eea
We thus encounter dissipative dynamics as in the case of spherical
symmetry: This is the equation of motion for a damped harmonic oscillator in one
dimension with time-dependent frequency and friction coefficient. The
solution of (\ref{m2}) with the initial conditions ${\tilde\pi}_x(0)=0$
and ${\tilde\pi}_y(0)=m\, v_y/2$ yields the kinematical momenta
\bea\label{eq:kinmomaxial}
2\,{\tilde\pi}_x(t)=m\,v_y
\sin\big(\mbox{$\frac{\omega\,v_z}{2}$}\, t^2\big) \qquad \mbox{and} \qquad
2\,{\tilde\pi}_y(t)=m\,v_y \cos\big(\mbox{$\frac{\omega\,
    v_z}{2}$}\, t^2\big) \ .
\eea
The classical trajectories starting from the origin are then given in terms of the
Fresnel integrals ${\rm S}(u)=\int_0^u\, \sin\big(\frac\pi2\, t^2\big)\, \dd t$
and ${\rm C}(u)= \int_0^u\, \cos\big(\frac\pi2\, t^2\big)\, \dd t$ as
\bea
x(t)= v_y \, \sqrt{\mbox{$\frac{\omega\,v_z}{\pi}$}}\ {\rm S}\Big(\sqrt{\mbox{$\frac{\pi}{\omega\,v_z}$}}\ t\Big)\qquad \mbox{and} \qquad
y(t)= v_y \, \sqrt{\mbox{$\frac{\omega\, v_z}{\pi}$}}\
{\rm C}\Big(\sqrt{\mbox{$\frac{\pi}{\omega\, v_z}$}}\ t\Big)\ .
\eea
A parameteric plot of the solution $\vec x(t) = \big(x(t),y(t),v_z\,t
\big)$ is displayed in Figure~\ref{fig:axialmotion}. 
The trajectory of the electric
charge is an Euler spiral along the straight line
$\big(v_y\, \sqrt{\frac{\omega\,v_z}{4\pi}}\,,\,v_y\, \sqrt{\frac{\omega\,v_z}{4\pi}}\,,\,v_z\,
t \big)$ in this case, to which the solution asymptotes at
$t\to\infty$. The particle moves with uniform velocity along the 
direction of the magnetic field and its motion
in the plane perpendicular to the field is confined. This is analogous to motion in a
uniform magnetic field $\vec B$, wherein \eqref{m2} is replaced by
the standard equation of motion for the one-dimensional harmonic
oscillator with the cyclotron frequency $\omega_{\rm cyc} = e\,|\vec
B|/m$ and the charged particle follows a helicoidal trajectory with
uniform velocity along the direction of $\vec B$.

\medskip

\medskip

Let us now consider the motion of a dyon in this magnetic background
by including an axial electric field $\vec E=(0,0,\varrho \,z)$, with $e\,\varrho>0$, which yields
a harmonic force corresponding to a confining potential
$V=\frac{e\,\varrho}2\, z^2$ in the $z$-direction. In this case the time
evolution of the axial position coordinate with the initial conditions
$z(0)=0$ and $\dot z(0)=v_z$ is given by $z(t)= v_z \sin(\varpi\,t)/\varpi$,
with oscillation frequency $\varpi=\sqrt{2\,e\,\varrho/m}$. The
differential equation for the kinematical momentum $2\,{\tilde\pi}_y$ then becomes
\bea
\ddot {\tilde\pi}_y-\varpi \cot(\varpi\,t) \, \dot
{\tilde\pi}_y+\mbox{$\frac{\omega^2\, v_z^2}{\varpi^2}$} \sin^2(\varpi\, t)\,  {\tilde\pi}_y=0 \ ,\label{m3}
\eea
which reduces to \eqref{m2} in the limit $\varrho=0$. For the special electric charge density $\varrho=m/2e$,
i.e.~$\varpi=1$, the appropriate choice of initial data gives the solutions
\bea
{\tilde\pi}_x(t)=\cos(\omega\,v_z \cos t) \qquad \mbox{and} \qquad
{\tilde\pi}_y(t)= \sin(\omega\,v_z \cos t) \ .
\eea
By electromagnetic duality, we should also set the magnetic
charge density to $\rho=m/2e$, i.e.~$\omega=1/2$. 
Since $\cos\big(\frac{v_z}2 \cos t\big)\geq\cos\big(\frac{v_z}2\big)>0$ if $v_z\leq3$,
the position coordinate $x(t)$ then increases monotonically with time, so that the motion
deconfines in the $(x,y)$-plane after confining the motion
along the $z$-direction. That is, the three-dimensional motion cannot
be completely confined, contrary to the expectations
of~\cite{Bojowald2015}. This suggests that the corresponding quantum Hamiltonian exhibits a
continuous energy spectrum. Below we shall investigate various
aspects of the quantum mechanics of electric charge in the monopole
background within our symplectic realisation.

Finally, we look at the dynamics of the auxiliary coordinates. With the physical solution \eqref{eq:kinmomaxial}, from \eqref{eom4} they evolve according to the equations of motion
\bea
\dot{\pi}_x &=& \omega\, v_z\, \big(t\, {\pi}_y+ \mbox{$\frac m2$} \,
\tilde y -\mbox{$\frac{m\,v_y}2$}\, t \cos(\mbox{$\frac{\omega\,
    v_z}{2}$}\, t^2) \big) \ , \nn\\[4pt]
\dot{\pi}_y &=& -\omega\, v_z\, \big(t\, {\pi}_x + \mbox{$\frac m2$}
\, \tilde x -\mbox{$\frac{m\,v_y}2$}\, t \sin(\mbox{$\frac{\omega\,
    v_z}{2}$}\, t^2) \big) \ , \nn\\[4pt]
\dot{\pi}_z &=& \mbox{$\frac{m\, \omega\, v_y}2$} \, \big(\cos(\mbox{$\frac{\omega\,
    v_z}{2}$}\, t^2) \, \tilde x - \sin(\mbox{$\frac{\omega\,
    v_z}{2}$}\, t^2) \, \tilde y\big) \ . \label{eq:auxeomaxial}
\eea
Thus the auxiliary degrees of freedom obey a complicated inhomogeneous system
of coupled differential equations, which we have not been able to
integrate; it would be interesting to
better understand the physical significance of the auxiliary
coordinates from these equations. This is in marked contrast to the
case of a uniform magnetic field $\vec B$, whereby \eqref{eom4} would
yield precisely the same Lorentz force law for the auxiliary variables, as expected since in that case the
constraints \eqref{c8} and \eqref{c9} can be consistently imposed.

This example illustrates that the symplectic realisation is not always necessary
for the integrability of the classical motion. However, it is
necessary for a proper formulation of geometric quantisation, i.e. for a description of the ``canonical'' quantum
mechanics. This is analogous to the situation of an
electric charge in the background of a Dirac monopole, wherein the classical equations of motion can be integrated without an action formalism, but for 
quantisation it is necessary to construct a suitable vector
potential in order to define a Hamiltonian.

\newsection{Quantum dynamics from symplectic realisation\label{sec:quantumdynamics}}

In this section we describe the quantisation of the dynamical system
with Poisson brackets (\ref{a5}) and Hamiltonian (\ref{Ham}). We shall
mostly ignore the electric background in our discussion and set the scalar
potential to $V=0$. In the Schr\"odinger polarisation, the quantum
Hilbert space $\CCH=L^2(\R^6)$ consists of square-integrable wavefunctions $\Psi$ (with respect to the standard Lebesgue measure) on the extended configuration space of the charged particle, which later on we will treat geometrically as sections of the trivial line bundle over $\R^6$ with connection $A_I$. On $\CCH$ we represent position operators $\widehat{x}{}^I$ as multipliers, $\big(\widehat{x}{}^I\Psi\big)(x)=x^I\, \Psi(x)$ with $x^I=(x^i,\tilde x^i)$, and the canonical momentum operators $\widehat{p}_I$ as derivatives, $\big(\widehat{p}_I\Psi\big)(x)=-\ii\hbar\,\p_I\Psi(x)$ with $\p_I=(\p_i,\tilde\p_i)$. Then the kinematical momentum operators 
\bea\label{eq:kinmomops}
\widehat{\pi}_I=\big(\widehat{\pi}_i,\widehat{\tilde\pi}_i\big) =-\ii\hbar\, \p_I-e\,A_I
\eea
define covariant differentiation on the trivial line bundle.

The corresponding quantum Hamiltonian is
\bea
\widehat{\mcH}=\mbox{$\frac{1}{m}$} \, \widehat{\pi}_I\, \eta^{IJ}\, \widehat{\pi}_J=\mbox{$\frac{1}{m}$} \, \big(\widehat{\pi}_i\, \widehat{\tilde\pi}{}^i+\widehat{\tilde\pi}_i\, \widehat{\pi}{}^i\big)=\mbox{$\frac2m$}\, \big(-\ii\hbar\, \vec{\nabla}-e\, \vec{A}\,\big)\cdot\big(-\ii\hbar\,\vec{\tilde\nabla}-e\,\vec{\tilde A}\,\big)\ .\label{c20}
\eea
The probability current $\CJ^I=(\CJ^i,\tilde{\CJ}{}^i)$ for a given state $\Psi$ is defined by
\bea
\CJ^I=\mbox{$\frac{1}{m}$} \, \big((\Psi^\ast\, \eta^{IJ}\, \widehat{p}_J\Psi-\Psi\, \eta^{IJ}\, \widehat{p}_J\Psi^\ast )-2\,e\, \eta^{IJ}\, A_J\, |\Psi|^2\big) \ .\label{c21}
\eea
Since the energy is conserved, as we have discussed in
Section~\ref{sec:integrability}, it suffices to consider stationary
states $\Psi$ which vary simply with a time-dependent phase and we can
study energy eigenvalues $\mcE$ of the quantum Hamiltonian \eqref{c20} via the time-independent
Schr\"odinger equation
\bea
\widehat{\mcH}\Psi = \mcE\, \Psi \ .
\label{eq:stationary}\eea
Due to \eqref{eq:stationary}, in a stationary state the probability current is conserved,
\bea
\p_I \CJ^I=0\ ,\label{c22}
\eea
and hence the quantum theory is also unitary.

This quantum theory is defined on a six-dimensional configuration space $(x^i, \tilde x^i)$. In particular, the probability current also has six components. Just as in the classical situation considered in Section~\ref{sec:Hamred}, we can try to eliminate the auxiliary degrees of freedom via quantum Hamiltonian reduction by imposing the constraints \eqref{c8} and \eqref{c9} (with $\zeta=\gamma=1$) at the quantum level, i.e. by restricting to the subspace $\CCH_{\rm phys}\subset\CCH$ of physical states $\Psi_{\rm phys}$ which are annihilated by the constraint operators: 
\bea
\widehat{\phi}{}^{\,i} \Psi_{\rm phys}=\big(\,\widehat{\tilde
  x}{}^{\,i}-\widehat{x}{}^{\,i} \big) \Psi_{\rm phys}=0 \qquad \mbox{and} \qquad \widehat{\psi}{}_i \Psi_{\rm phys}=\big(\,\widehat{\tilde\pi}{}_i-\widehat{\pi}{}_i\big) \Psi_{\rm phys}=0 \ .\label{c23}
\eea
The relevant commutator to analyse is given by
\bea
\big[\,\widehat{\tilde\pi}_i,\widehat{\pi}_j\big]=\big[\widehat{\pi}_i,\widehat{\tilde\pi}_j\big]=\mbox{$\frac{\ii
    \hbar\, e}2$}\, \varepsilon_{ijk}\, \widehat{B}{}^k \ ,
\eea
where $\widehat{B}{}^k$ is the multiplier by the magnetic field $\vec B(\vec x\,)$.

If the magnetic field $\vec B$ is divergenceless, i.e.  $\vec B=\vec\nabla\times\vec a$ everywhere, then the vector potential $A_I(x^I)$ can be defined as 
\bea
\vec A(x^I)=\mbox{$\frac12$}\, \big(\vec a(\vec x\,)-(\vec x\cdot
\vec\nabla)\vec a(\vec x\,)-\vec{\tilde x}\times\vec B(\vec x\,)\big)
\qquad \mbox{and} \qquad \vec{\tilde A}(x^I)=\vec0\ .\label{c24}
\eea 
With this choice, the effective quantum theory after resolving the constraints
\eqref{c23} coincides with the standard quantum mechanical
description of a charged particle in the magnetic field with the Hamiltonian
\bea
\widehat{\mcH}_{\rm eff}=\mbox{$\frac{1}{2m}$}\, \widehat{\pi}_i\,
\widehat{\pi}^{\,i} \ ,\label{c25}
\eea 
where 
\bea
\big[\widehat{\pi}_i,\widehat{\pi}_j\big]=\ii \hbar\,e\,
\varepsilon_{ijk}\, \widehat{B}{}_{\rm eff}^k=\ii \hbar\, e\, \big(\widehat{\p_ia_j} -\widehat{\p_ja_i} \big)\ .\label{c26}
\eea

In situations where
the vector field $\vec a(\vec x\,)$ is not globally defined on
$\mathbb{R}^3$ but only in specified compact regions, like in the case of
the Dirac monopole, one can excise the support of the magnetic charge
distribution $\vec\nabla\cdot\vec B$ from the configuration space and
take the wavefunctions to be sections of a corresponding non-trivial
line bundle over the excised space of degree $n\in\Z$ given by the Dirac charge quantisation
condition \eqref{eq:Diracquant}~\cite{Wu1976}. In the standard treatments one needs
to restrict the domains of quantum operators to wavefunctions which
vanish sufficiently fast on the Dirac string, whereas in our approach we can
simply consider wavefunctions in $\CCH_{\rm
  phys}=L^2(\R^3)$ that vanish at the locus of the magnetic charge
distribution, which provides a suitable extension of the
effective Hamiltonian \eqref{c25} to an essentially self-adjoint operator on~$\CCH_{\rm
  phys}$~\cite{BL}. It is known that a Dirac monopole and an electric
charge do not form a bound state, whereas dyonic bound states are
possible~\cite{Zwanziger1968}.

However, if $\vec\nabla\cdot\vec B\neq0$ everywhere then a vector potential
$\vec a(\vec x\,)$ does not exist even locally and the effective magnetic
field in
(\ref{c26}) does not account for the contribution from magnetic
sources. In particular, for the spherically symmetric magnetic field \eqref{eq:Brot} it simply vanishes,
$\vec B_{\rm eff}=\vec0$, and thus in
this case the constraints (\ref{c23}) lead to a free particle quantum theory
without any interaction with the magnetic field.

To understand better the quantum theory described by the Hamiltonian
(\ref{c20}), we first observe that it is an unbounded operator on $\CCH$. It is convenient to represent it as a {\it difference} of two Hamiltonians which are each bounded from below as
\bea
\widehat{\mcH}=\widehat{\mcH}_+-\widehat{\mcH}_-:=\mbox{$\frac {1}{ m}$} \, \widehat{ \pi}{}_{i\,+}\,\widehat{\pi}^{\,i}_+ -\mbox{$\frac {1}{ m}$} \, \widehat{\pi}{}_{i\,-}\,\widehat{\pi}^{\,i}_- \ ,\label{ll1}
\eea
where
\bea
\widehat{\pi}{}_{i\,\pm}=\mbox{$\frac{1}{\sqrt 2}$} \,
\big(\widehat{\pi}_i\pm\widehat{\tilde\pi}_i\big) \ . \label{ll2}
\eea
The pairs of kinematical momentum operators \eqref{ll2} do not commute in
general,
\begin{equation}
\big[\widehat{\pi}{}_{i\,+},\widehat{\pi}{}_{j\,-}\big] =\mbox{$\frac{\ii
    \hbar\, e}2$}\,\big(\varepsilon_{ijk}\, \widehat{\p_l B^k}
-\varepsilon_{ijl}\, \widehat{\p_k B^k} \big)\, \widehat{\tilde x}{}^{\,l} \ ,
\end{equation}
and consequently neither do the Hamiltonians $\widehat{\mcH}_+$ and
$\widehat{\mcH}_-$ unless the magnetic field is constant.
Let us begin by considering some typical examples which illustrate
how the imposition of the constraints (\ref{c23}) recovers well-known
results, before moving on to our main examples of interest with smooth
distributions of magnetic charge.

\bigskip

\noindent
{\bf Free particle. \ } 
As a warmup, let us see how to reproduce free particle quantum states in the
absence of a magnetic field, $A_I=0$. The Schr\"odinger equation
\eqref{eq:stationary} in this case is
\bea
-\frac{2\hbar^2}{m} \, \vec{\nabla}\cdot\vec{\tilde\nabla}
\Psi(\vec x,\vec{\tilde x}\,)=\mcE\, \Psi(\vec x,\vec{\tilde x}\, )\ .
\eea
The eigenfunctions are the plane waves
\bea
\Psi_{\vec k,\vec{\tilde k}}(\vec x,\vec{\tilde x}\,)=\e^{-\frac{\ii}{2\hbar}\, (\vec
  k\cdot\vec x+\vec{\tilde k}\cdot\vec{\tilde x}\, )}\ ,
\eea
with eigenvalues
\bea
\mcE_{\vec k,\vec{\tilde k}}=\frac{\vec k\cdot\vec{\tilde k}}{2m} \ .
\eea
The physical state conditions \eqref{c23} then force $\vec x=\vec{\tilde
  x}$ and $\vec k=\vec{\tilde k}$, yielding the expected
free particle plane waves and kinetic energy spectrum
\bea
\Psi_{\rm phys}(\vec x\,) = \Psi_{\vec k,\vec k}(\vec x,\vec x\,) =\e^{-\frac\ii\hbar\,
  \vec k\cdot \vec x} \qquad \mbox{and} \qquad E_{\vec k}=\mcE_{\vec k,\vec
  k}=\frac{\vec k^{\,2}}{2m} \ .
\eea

\bigskip

\noindent
{\bf Landau levels. \ } 
Consider the Landau problem, i.e. the motion of an electric charge in a constant and uniform magnetic field $\vec B$, which by a suitable choice of coordinates we can take to lie along the $z$-axis, $\vec B=(0,0,B)$. In this case
\bea
\vec A(x^I)=\mbox{$\frac B2$}\, (-\tilde y,\tilde x, 0) \qquad \mbox{and} \qquad \vec{\tilde A}(x^I)=\vec0\ ,\label{ll3}
\eea 
and the only non-vanishing commutators between the covariant momentum operators are given by
\bea
\big[\widehat{\pi}{}_{x\,\pm},\widehat{\pi}{}_{y\,\pm}\big]=\pm \ii \hbar \, e\, B \ .\label{ll4}
\eea
In particular, the axial momentum operators $\widehat{p}_z$ and
$\widehat{\tilde p}_z$ commute with all other momentum operators, and
so the quantum states in the direction of $\vec B$ decouple into free
particle states which can be solved for along the same lines as
above. Henceforth we therefore consider only the planar quantum states and,
assuming $e\,B>0$, we introduce creation and annihilation operators as 
\bea
\widehat{a}_\pm=\mbox{$\frac{1}{\sqrt{2 \, \hbar\,e\, B}}$}\,  \big(\widehat{\pi}{}_{x\,\pm}+\ii\widehat{\pi}{}_{y\,\pm}\big) \qquad \mbox{and} \qquad \widehat{a}{}^\dagger_\pm=\mbox{$\frac{1}{\sqrt{2 \, \hbar\,e\, B}}$}\,  \big(\widehat{\pi}{}_{x\,\pm}-\ii\widehat{\pi}{}_{y\,\pm}\big) \ . \label{ll5}
\eea
One easily checks
\bea
\big[\widehat{a}_\pm,\widehat{a}{}^\dagger_\pm\big]= 1\ ,\label{ll6}
\eea
while all other commutators vanish. 

In terms of the operators (\ref{ll5}) the Hamiltonian (\ref{ll1}) can be written as 
\bea
\widehat{\mcH}=\widehat{\mcH}_+-\widehat{\mcH}_-:= \hbar \, \omega_{\rm cyc}\, \big( \widehat{a}{}^\dagger_+\, \widehat{a}_+-\widehat{a}{}^\dagger_-\, \widehat{a}_--1\big)\ ,\label{ll7}
\eea
where
\bea
\omega_{\rm cyc}=\frac{e\,B}{m} 
\eea
is the cyclotron frequency. This Hamiltonian is unbounded, but it can
be decomposed into the difference of two Hamiltonians
$\widehat{\mcH}_\pm$ of harmonic oscillator type which are bounded from
below and commute, $\big[\widehat{\mcH}_+, \widehat{\mcH}_-\big]=0$. As
discussed in Appendix~\ref{app:dho}, the simultaneous eigenvalues of
$\widehat{a}{}^\dagger_\pm\, \widehat{a}_\pm$ are the integers
$n_\pm\in\N_0$ such that the eigenvalues of (\ref{ll7}) are
\bea
\mcE_{n_+,n_-}= \hbar\,\omega_{\rm cyc}\, (n_+-n_--1)
\eea
with
corresponding eigenstates $|n_+,n_-\rangle$ in the standard number
basis for the two-particle bosonic Fock space $\CCF$. Using the
definition of the annihilation operator $\widehat{a}_-$ from
\eqref{ll5}, the physical state constraints \eqref{c23} read
$\widehat{a}_-\Psi_{\rm phys}=0$, which implies that the
$\widehat{a}_-$-oscillator must be kept in its ground state for which
$\widehat{a}_-|n_+,0\rangle=0$. Then the standard harmonic oscillator
spectrum $E_{n}=\mcE_{n,0}$, and hence the Landau levels of the electric charge, emerge. This is the same as the known constraint from the quantum theory of dissipative dynamics~\cite{Feshbach,Dekker,Celeghini,tHooft}, see Appendix~\ref{app:dho}.

\bigskip

\noindent
{\bf Axial magnetic fields. \ } 
The natural extension of the Landau problem considered above is to the motion of an electric charge in a constant and uniform magnetic charge distribution which sources an axial magnetic field \eqref{eq:Baxial}. In this case the vector potential is given by \eqref{eq:Aaxialmod} and the Hamiltonian reads
\bea
\widehat{\mcH}=\mbox{$\frac {1}{m}$}\,
\big(\widehat{\pi}{}^2_{x\,+}+\widehat{\pi}{}^2_{y\,+}\big)-\mbox{$\frac
  {1}{m}$}\,
\big(\widehat{\pi}{}^2_{x\,-}+\widehat{\pi}{}^2_{y\,-}\big) +
\mbox{$\frac2m$}\, \widehat{p}_{z}\, \widehat{\tilde
  p}_{ z}\ ,
\eea
with the algebra of non-vanishing commutation relations among momentum operators given by
\bea
\big[\widehat{\pi}_{x\,\pm},\widehat{\pi}_{y\,\pm}\big] &=& \pm \,
\mbox{$\frac{\ii \hbar}4$} \, e \, \rho \, \widehat{z} \ , \nn \\[4pt]
\big[\widehat{\pi}_{x\,\pm},\widehat{p}_{z}\big] &=&
\pm\, \mbox{$\frac{\ii \hbar}{4\,\sqrt2}$}\, e\, \rho \,
\big(\,\widehat{y}-2\,\widehat{\tilde y}\,\big) \ , \nn \\[4pt]
\big[\widehat{\pi}_{y\,\pm},\widehat{p}_{z}\big] &=& \pm\, \mbox{$\frac{\ii \hbar}{4\,\sqrt2}$}\, e\, \rho \, \big(\,2\,\widehat{\tilde x}-\widehat{x}\,\big) \ . 
\eea
In particular $\big[\widehat{\mcH},\widehat{\tilde p}_z\big]=0$, so
the quantum states form representations of the translation group generated by $\widehat{p}_z$, which are superpositions of the simultaneous eigenstates of the axial momentum operator given by
\bea
\Psi_{\tilde p_z}\big(\vec x,\vec{\tilde x}\,\big) = \e^{\ii\tilde p_z\,\tilde
  z/\hbar} \ \Psi(x,y,z,\tilde x,\tilde y) \qquad 
\mbox{with} \quad \widehat{\tilde p}_z\Psi_{\tilde p_z} = \tilde
    p_z \, \Psi_{\tilde p_z} \ . 
\eea
This defines a decomposition of the quantum Hilbert space $\CCH$ into
a direct integral $\int^\oplus_{\tilde p_z\in\real}\, \CCH_{\tilde
  p_z} := L^2(\real,\CCH)$, the square-integrable sections of the
state space viewed as a (trivial) Hilbert bundle over the line $\real$ of axial momenta. The Schr\"odinger equation \eqref{eq:stationary} in the fiber subspace over $\tilde
p_z\in\real$ is equivalent to $\widehat{\mcH}_{\tilde p_z}\Psi = \mcE\,
\Psi$, with the restriction of the Hamiltonian $\widehat{\mcH}_{\tilde p_z}:= \widehat{\mcH}\big|_{\CCH_{\tilde p_z}}$ given by
\bea\label{eq:Hamtildepz}
\widehat{\mcH}_{\tilde p_z} =\mbox{$\frac {1}{m}$}\,
\big(\widehat{\pi}{}^2_{x\,+}+\widehat{\pi}{}^2_{y\,+}\big)-\mbox{$\frac
  {1}{m}$}\,
\big(\widehat{\pi}{}^2_{x\,-}+\widehat{\pi}{}^2_{y\,-}\big) +
\mbox{$\frac{2}m$}\, \tilde p_z \ \widehat{p}_{z} \ .
\eea

Let us now introduce the frequency
\bea
\omega = \frac{e\, \rho}m
\eea
which appears in the classical solution of
Section~\ref{sec:integrability}, and we assume again that
$e\,\rho>0$. We further introduce the
``creation'' and ``annihilation'' operators
\bea
\widehat{a}_\pm = \sqrt{\mbox{$\frac{2}{\hbar\,m\,\omega}$}} \
  \big(\widehat{\pi}_{x\,\pm}+\ii\widehat{\pi}_{y\,\pm}\big) \qquad
  \mbox{and} \qquad \widehat{a}{}^\dag_\pm = \sqrt{\mbox{$\frac{2}{\hbar\,m\,\omega}$}} \
  \big(\widehat{\pi}_{x\,\pm}-\ii\widehat{\pi}_{y\,\pm}\big) \ ,
\eea
with the non-vanishing commutation relations
\bea\label{eq:hatzoscalg}
\big[\widehat{a}_\pm,\widehat{a}{}^\dag_\pm\big] = \widehat{z} \ .
\eea
In particular, the axial position operator $\widehat{z}$ is central in this
``oscillator'' algebra: $\big[\widehat{a}_\pm,\widehat{z}\big] = 0$ and
$\big[\widehat{a}{}^\dag_\pm,\widehat{z}\big]=0$. One easily checks the further non-vanishing
commutators
\bea\label{eq:hatapzcomm}
\big[\widehat{a}_\pm,\widehat{p}_z\big] = \mp\,
\sqrt{\mbox{$\frac{\hbar}{m\,\omega}$}} \ \mbox{$\frac e2$} \ 
\widehat{w} \qquad \mbox{and} \qquad
\big[\widehat{a}{}^\dag_\pm,\widehat{p}_z\big] = \pm\,
\sqrt{\mbox{$\frac{\hbar}{m\,\omega}$}} \ \mbox{$\frac e2$} \ 
\widehat{w}{}^\dag \ ,
\eea
where
\bea
\widehat{w} = \big(\,\widehat{\tilde x}+\ii\widehat{\tilde y}\,\big)
-\mbox{$\frac12$}\, \big(\widehat{x}+\ii\widehat{y}\big) \qquad
\mbox{and} \qquad \widehat{w}{}^\dag = \big(\,\widehat{\tilde
  x}-\ii\widehat{\tilde y}\, \big)
-\mbox{$\frac12$}\, \big(\widehat{x}-\ii\widehat{y}\big) \ ,
\eea
and of course one has the canonical commutator
\bea\label{eq:hatzpzcomm}
\big[\widehat{z}\,,\,\widehat{p}_z\big] = \ii\hbar \ .
\eea
The Hamiltonian \eqref{eq:Hamtildepz} then becomes
\bea\label{eq:axialosc}
\widehat{\mcH}_{\tilde p_z} = \mbox{$\frac\hbar2$} \, \omega\,
\big(\widehat{a}{}^\dag_+\, \widehat{a}_+ - \widehat{a}{}^\dag_-\,
\widehat{a}_- - \widehat{z}\big) + \mbox{$\frac{2}m$}\, \tilde p_z \
\widehat{p}_z \ . 
\eea

Naively, this Hamiltonian resembles that of the Landau problem considered above, in that it decomposes into a free particle Hamiltonian in the axial direction plus a doubled ``oscillator'' system in the planar directions. However, due to \eqref{eq:hatapzcomm} and \eqref{eq:hatzpzcomm}, these two components do not commute, and moreover the planar ``oscillator'' depends explicitly on the axial position operator through \eqref{eq:hatzoscalg}. This coupling between the planar and axial momentum operators hinders a complete analytic solution of the Schr\"odinger equation, in contrast to the classical dynamics from Section~\ref{sec:integrability} where the free motion in the axial direction effectively reduces the problem to planar motion in a time-dependent magnetic
field which we were able to integrate. Note that on physical states
$\Psi_{\rm phys}$ satisfying the analogue of the constraint equations
\eqref{c23} with $\zeta=\frac12$ in \eqref{c8} one has $\widehat{w} \Psi_{\rm phys}=0=\widehat{w}{}^\dag
\Psi_{\rm phys}$, so that the planar and axial Hamiltonians commute on $\CCH_{\rm phys}$, but now additionally $\widehat{a}_+\Psi_{\rm phys}=0=\widehat{a}_-\Psi_{\rm phys}$ from \eqref{c23} so that the planar quantum dynamics trivialises and the free particle axial states follow as before.

For states with vanishing axial momentum $\tilde p_z=0$, the spectrum of the Hamiltonian
\bea\label{eq:H0}
\widehat{\mcH}_{0} = \mbox{$\frac\hbar2$} \, \omega\,
\big(\widehat{a}{}^\dag_+\, \widehat{a}_+ - \widehat{a}{}^\dag_-\,
\widehat{a}_- - \widehat{z}\big)
\eea
is readily obtained: Via Fourier transformation of the Schr\"odinger
polarisation, we can represent the axial position operator as the
derivative $\widehat{z} = \ii\hbar\, \frac\partial{\partial p_z}$ so that the subspace $\CCH_0=\CCF\otimes L^2(\real)$ is spanned by the eigenstates
\bea
\Psi^{(0)}_{n_+,n_-;z}(p_z) = |n_+,n_-\rangle \otimes\psi_z(p_z)
\eea
of \eqref{eq:H0}, where $\psi_z(p_z) = \e^{-\ii z\,p_z/\hbar}$ are the eigenstates of $\widehat{z}$ with axial position eigenvalue $z\in\real$. The correponding energy eigenvalues are
\bea\label{eq:cE0}
\mcE^{(0)}_{n_+,n_-;z} = \mbox{$\frac\hbar2$} \, \omega \, z \, \big(n_+-n_--1\big) \ ,
\eea
which is the spectrum of a doubled harmonic oscillator with an axial
position-dependent frequency. For small momenta, via a suitable
(length) regularisation of the inner product on $L^2(\real)$ it is
easy to see that the first order correction to these
energies due to the perturbation by the axial momentum operator in
\eqref{eq:axialosc} vanishes, so that \eqref{eq:cE0} represents the
energy of the system up to order $O\big(\tilde p{}^2_z\big)$. 
Thus the quantum dynamics of the electric charge in an axial magnetic field exhibits a continuous energy spectrum, and by our calculations from Section~\ref{sec:integrability} we do not expect the situation to change by inclusion of a corresponding axial electric field.

It would be interesting to find the exact spectrum of the Hamiltonian \eqref{eq:axialosc}, but we will content ourselves here with the approximate solution \eqref{eq:cE0}. The situation is of course much more complicated for the spherically symmetric magnetic field \eqref{eq:Brot}, whose dyonic classical Hamiltonian is given by \eqref{eq:Hspherdyon}, due to the fact that fewer integrals of motion exist in that case. In Sections~\ref{sec:magnetictranslations} and~\ref{sec:NAQM} below we shall discuss some of the features of the charged particle wavefunctions in the spherically symmetric case.

\newsection{Extended magnetic translations and two-cocycles\label{sec:magnetictranslations}}

One of the most interesting aspects of the quantum dynamics of electric
charge in magnetic backgrounds is the physical and mathematical
structure of the magnetic translation group~\cite{Zak1964}. For source-free magnetic
fields, the electron wavefunctions carry a (weak) projective representation
of the translation group $\real^3$ whose two-cocycle is defined by the
magnetic flux. On the other hand, for magnetic fields sourced by
monopole distributions, the representation of the translation group is
obstructed by an anomalous three-cocycle defined by the magnetic charge, which
encodes a ``nonassociative representation'' in the sense that the
parallel transports implementing the translations do not associate~\cite{Jackiw1985};
in this case one cannot assign operators to the translation generators
which act on a separable Hilbert space and one is forced to deal with
other methods of quantisation, such as the phase space formulation of
nonassociative quantum mechanics~\cite{Mylonas2013,SzaboISQS}, or the action of parallel transports on a 2-Hilbert space which generates higher
projective representations~\cite{Bunkinprep}. In this section we
wish to see how these obstructing three-cocycles are captured within
the associative framework of our symplectic realisation, following the
treatment of magnetic translations
from~\cite{Hannabuss2017,Soloviev2017} which we adapt to our situation.

The key feature of the symplectic realisation is the existence of a globally defined vector potential \eqref{eq:extendedA}, which we interpret geometrically as a connection on the trivial line bundle over $\real^6$. Gauge invariance of the Schr\"odinger equation \eqref{eq:stationary} dictates that a gauge transformation $A_I\mapsto A_I+\partial_I\chi$ is accompanied by a corresponding phase transformation $\Psi\mapsto \e^{\ii e\,\chi}\, \Psi$ of the electron wavefunctions $\Psi\in\CCH$. In the presence of the
magnetic field $\vec B(\vec x\,)$, the translation generators $\partial_I$ on
the extended configuration space $\real^6$ are modified to the
kinematical momentum operators \eqref{eq:kinmomops}, and hence we must
extend the natural operators which generate translations by vectors $r=(\vec
r,\vec{\tilde r}\,)\in\real^6$ on the quantum Hilbert space
$\CCH=L^2(\R^6)$ to magnetic translation operators $\Tw(r)$ which act on
wavefunctions $\Psi\in\CCH$ at a point $x\in\real^6$ by parallel transport along the line connecting $x$ to $x+r$:
\bea\label{eq:partrans}
\big(\Tw(r)\Psi\big)(x) = \exp\Big(\,\frac{\ii e}{\hbar}\, \int_0^1\,
\dd t\ r^I\,A_I(x+t\,r)\,\Big)\,\Psi(x+r) \ .
\eea
This defines a one-cochain of the translation group in six dimensions.

The operator $\Tw(r)\, \Tw(s)\, \Tw(r+s)^{-1}$ performs the parallel
transport of the wavefunction $\Psi(x)$ around the loop forming 
the boundary $\partial\triangle(x;r,s)$ of the triangle $\triangle(x;r,s)$ based at $x\in\real^6$
and spanned by the translation vectors $r,s$. In terms of the
one-form $A=A_I\, \dd x^I=\vec A(\vec x,\vec{\tilde x}\,)\cdot \dd
\vec x$ on the extended configuration space whose components are given by
\eqref{eq:extendedA}, this has the effect of multiplying $\Psi(x)$ by
the Wilson loop  $W(x;r,s)$ of the gauge field $\frac e\hbar\,A$ around
$\partial\triangle(x;r,s)$. We then obtain the relations 
\bea\label{eq:projrep}
\Tw(r)\,\Tw(s) = \widehat{\Omega}(r,s)\, \Tw(r+s) \ ,
\eea
where the mutually commuting quantum operators $\widehat{\Omega}(r,s)$ for
$r,s\in\real^6$ are the
multipliers 
\bea\label{eq:projmultiplier}
\big(\widehat{\Omega}(r,s)\Psi\big)(x) := W(x;r,s)\, \Psi(x) \ .
\eea
The phase factor $W(x;r,s)$ is the coboundary of the one-cochain
defined by the parallel transport \eqref{eq:partrans} which reads as
\bea\label{eq:projphase}
W(x;r,s) := \exp\Big(\,\frac{\ii e}\hbar\,
\oint_{\partial\triangle(x;r,s)}\, A\,\Big) = \exp\Big(\,\frac{\ii
  e}\hbar\, \int_{\triangle(x;r,s)}\, F\,\Big) \ ,
\eea
where $F=\dd A$ is the field strength of $A$ whose components are given by \eqref{eq:piIpijcurl}
and we have used Stokes' theorem.\footnote{Formally we may regard
$\widehat{\Omega}(r,s):= W(\,\widehat x\,;r,s)$.} 

By construction the extended
magnetic translation operators associate,
\bea
\big(\Tw(r)\,\Tw(s)\big)\,\Tw(u) = \Tw(r)\, \big(\Tw(s)\,\Tw(u)\big) \ ,
\eea
which implies that the multipliers of \eqref{eq:projrep} satisfy the
two-cocycle condition
\bea\label{eq:2cocycle}
\widehat{\Omega}(r,s)\, \widehat{\Omega}(r+s,u) = {}^r{}\widehat{\Omega}(s,u)\,\widehat{\Omega}(r,s+u) \ ,
\eea
where ${}^r{}\widehat{\Omega}(s,u):= \Tw(r)\, \widehat{\Omega}(s,u)\, \Tw(r)^{-1}$ is the
multiplier
\bea
\big({}^r{}\widehat{\Omega}(s,u)\Psi\big)(x) := W(x+r;s,u)\, \Psi(x) \ .
\eea
The relations \eqref{eq:projrep} and \eqref{eq:2cocycle} imply that the map $r\mapsto\Tw(r)$ 
defines a weak projective representation of the
translation group $\real^6$ on the quantum Hilbert space of states
$\CCH$, where by ``weak'' we mean that the projective phase is a multiplier by
\eqref{eq:projphase} which has a non-trivial dependence on position
coordinates $x\in\real^6$~\cite{Soloviev2017}.

Using the Poisson algebra \eqref{a5}, it is easy to compute the phase
factor \eqref{eq:projphase} explicitly in terms of surface integrals
over triangles in the extended configuration space to get
\bea
W(x;r,s) &=& \exp\bigg[\,\frac{\ii e}\hbar\, \Big(\, \frac12\,
\int_{\triangle(\vec x;\vec r,\vec{\tilde s}\,)}\, \vec B\cdot \dd\vec S + \frac12\,
\int_{\triangle(\vec x;\vec{\tilde r},\vec{ s}\,)}\, \vec B\cdot
\dd\vec S \nonumber \\ && \qquad \qquad \qquad +\, \int_{\triangle(\vec x;\vec r,\vec s\,)}\,
\big(\vec\nabla\cdot\vec B\,\big)\, \vec{\tilde x}\cdot\dd\vec
S\,\Big)\,\bigg] \ .
\eea
The third integration can be expressed as a volume integral over
the tetrahedron $\triangle(\vec x;\vec{\tilde x},\vec r,\vec s\,)$
based at $\vec x\in\real^3$ and spanned by the vectors $\vec{\tilde
  x},\vec r,\vec s$. Altogether we then find
\bea\label{eq:phialtogether}
W(x;r,s) = \exp\bigg[\,\frac{\ii e}\hbar\, \Big(\, \frac12\,
\int_{\triangle(\vec x;\vec r,\vec{\tilde s}\,)\cup \triangle(\vec x;\vec{\tilde r},\vec{ s}\,)}\, \vec B\cdot \dd\vec S + \int_{\triangle(\vec x;\vec{\tilde x},\vec r,\vec s\,)}\,
\vec\nabla\cdot\vec B \ \dd V\,\Big)\,\bigg] \ .
\eea
The phase integrals in \eqref{eq:phialtogether}, which are each
defined in terms of auxiliary coordinates, combine to give a hybrid of
the usual magnetic flux two-cocycle in the source-free case and the
magnetic charge three-cocycle in the presence of monopoles, in such a
way so that $W(x;r,s)$ itself defines a two-cocycle of the extended translation group $\real^6$. Let us look at a few special cases to understand this structure more thoroughly.

We start with the source-free case, $\vec\nabla\cdot\vec B=0$, so that the second integral in \eqref{eq:phialtogether} vanishes. As discussed in Section~\ref{sec:quantumdynamics}, in this instance one can consistently implement quantum Hamiltonian reduction through the constraints \eqref{c23}, and by restricting the action of the multipliers \eqref{eq:projmultiplier} to physical states $\Psi_{\rm phys}\in\CCH_{\rm phys}\subset\CCH$, one can identify physical and auxiliary translations and coordinates in \eqref{eq:phialtogether} to get
\beq
W_0(\vec x;\vec r,\vec s\,) = \exp\Big(\,\frac{\ii e}\hbar\,
\int_{\triangle(\vec x;\vec r,\vec{s}\,)}\, \vec B\cdot \dd\vec S\,\Big) \ .
\eeq
Thus in this case we recover the standard two-cocycle of the
anticipated (weak) projective representation of the translation group $\real^3$~\cite{Zak1964}.

Next, for the Dirac monopole field \eqref{eq:BDirac}, by restricting to wavefunctions which vanish at the origin of $\real^3$ as discussed in Section~\ref{sec:quantumdynamics}, one may again impose the constraints \eqref{c23}, and hence identify physical and auxiliary variables in \eqref{eq:phialtogether}. The second integral now computes the magnetic charge enclosed by the tetrahedron $\triangle(\vec x,\vec r,\vec{s}\,)$, whose contribution to the phase is unity due to the Dirac charge quantisation condition \eqref{eq:Diracquant}. In this way we reproduce the result of~\cite{Jackiw1985} for the projective two-cocycle phase
\bea
W_{\rm D}(\vec x;\vec r,\vec s\,) = \exp\Big(\,\frac{\ii e}\hbar\,
\int_{\triangle(\vec x;\vec r,\vec{s}\,)}\, \vec B_{\rm D}\cdot \dd\vec S\,\Big)
\eea
generated by the Dirac monopole.

Finally, let us consider the case of a constant magnetic charge distribution with spherically symmetric magnetic field \eqref{eq:Brot}. Contrary to the previous two cases, in this instance one cannot impose quantum Hamiltonian reduction to eliminate the auxiliary variables, and explicit computation of \eqref{eq:phialtogether} yields
\bea\label{eq:phirot}
W_{\rm spher}(x;r,s) = \exp\bigg[\,\frac{\ii e\,\rho}\hbar\, \Big(\, \frac16\, \big[\vec{\tilde r},\vec s,\vec x\, \big] + \frac16\, \big[\vec r,\vec{\tilde s},\vec x\, \big] + \big[\vec r,\vec s,\vec{\tilde x}\, \big]\,\Big)\,\bigg] \ ,
\eea
where the triple scalar product $\big[\vec r,\vec s,\vec{\tilde x}\, \big]:= \vec r\cdot\big(\vec s\times \vec{\tilde x} \, \big)$ is the volume of the tetrahedron $\triangle\big(\vec r,\vec s,\vec{\tilde x}\, \big)$ in the extended configuration space $\real^6$. The third phase contribution in \eqref{eq:phirot} is the analogue of the three-cocycle of the translation group $\real^3$ which is calculated in nonassociative quantum mechanics~\cite{Mylonas2013,SzaboISQS}; in the associative symplectic realisation, it is defined here by inserting an auxiliary position vector $\vec{\tilde x}$ into the third argument of the three-cocycle. Again, all three phase contributions together in \eqref{eq:phirot} ensure a weak projective phase that defines a two-cocycle of the extended translation group $\real^6$.

In nonassociative quantum mechanics~\cite{Mylonas2013,SzaboISQS}, the non-trivial three-cocycle in the case of uniform magnetic charge density $\rho$ has profound physical consequences on the quantum system: It leads to a quantised momentum space with a quantum of minimal volume $\frac12\,\hbar^2\,\rho$. In canonical (associative) quantum mechanics such volume quantisation is not observable, because there is no non-trivial volume operator. However, minimal areas are observable, such as the phase space Planck cell quantum $\hbar$, and for the present discussion the pertinent operator measuring area uncertainties in the extended momentum space is given by setting $\widehat{\bar\pi}_{\vec r}:= r^i\, \widehat{\bar\pi}_i$ for $\vec r\in\real^3$ and defining
\bea
\widehat{A}_{\vec r,\vec s} := {\rm Im}\,\big[\,\widehat{\bar\pi}_{\vec r}\,,\,\widehat{\bar\pi}_{\vec s}\,\big] \ .
\eea
The idea behind this definition is that the vector product of two
vectors $\vec r, \vec s$ from the physical subspace is a vector in the
extended configuration space, and so the operator $\widehat{A}_{\vec
  r,\vec s}$ measures a physical volume in the extended space. Indeed,
using the Poisson algebra \eqref{PB1}, we may compute the expectation
value of this oriented area operator in any state $\Psi\in\CCH$ (with the
standard $L^2$-inner product) to get
\bea\label{eq:minimalvolume}
\big\langle\Psi\big| \,\widehat{A}_{\vec r,\vec s}\, \big|\Psi\big\rangle
= \mbox{$\frac{\hbar\,e\,\rho}3$} \, \big[\vec r,\vec
s,\delta_\Psi\vec x\, \big] \qquad \mbox{with} \quad \delta_\Psi x^i:= \big\langle\Psi\big| \,\widehat{\tilde
  x}{}^{\,i}-\widehat{x}{}^{\,i} \big|\Psi\big\rangle \ .
\eea
We have seen in this case that the quantum dynamics in magnetic charge
backgrounds is not consistent with the physical state conditions \eqref{c23}, and hence the quantum tetrahedral volume computed by \eqref{eq:minimalvolume} is generically non-zero. This is the sense in which our associative formalism realises the characteristic minimal volumes. In Section~\ref{sec:NAQM} below we shall demonstrate more precisely the correspondence between the symplectic realisation and nonassociative quantum mechanics.

\newsection{Nonassociative quantum
  mechanics\label{sec:NAQM}}

In this final section we shall conclude with somewhat more formal
developments. One of our motivations for the present study was to
understand the somewhat mysterious composition
product that underlies the associative algebra of observables in
nonassociative quantum mechanics~\cite{Mylonas2013}, regarded as a
nonassociative deformation of the standard phase space formulation of quantum mechanics~\cite{Zachos2001}, and provides an
associative realisation of nonassociative star products. Deformation
quantisation of the magnetic monopole algebra \eqref{a1} was
originally carried out via explicit
construction of a nonassociative star product in~\cite{Mylonas2012}
(see also~\cite{BL,Kupriyanov2015}), and cast into a quasi-Hopf algebraic
framework in~\cite{Mylonas2013}. We shall demonstrate how the associative
realisation of nonassociative quantum mechanics in terms of
composition products from~\cite{Mylonas2013} can be realised explicitly in terms of an
algebra of differential operators on phase space, and then show that
this is identical to the quantum algebra given by the symplectic
realisation of the underlying twisted Poisson structure. 

\bigskip

\noindent
{\bf Nonassociative star product. \ } 
For definiteness, throughout this section we work explicitly with a
uniform monopole density of strength $\rho$ in three dimensions
and the spherically symmetric magnetic field
\eqref{eq:Brot}; the analysis can be generalised to non-constant
magnetic charge distributions, at least perturbatively.\footnote{Associative star products quantising the Poisson brackets corresponding to the field of a Dirac monopole are discussed e.g. in~\cite{Carinena:2009ug,Soloviev2017}.} For notational ease, we write the corresponding magnetic monopole algebra
\eqref{a1} collectively in terms of a twisted Poisson bivector
$\theta$ on phase space $\CP$ as
\bea
\{x^a,x^b\}=\theta^{ab}(x)=
\bigg(\begin{matrix}
0&\delta^i{}_j\\
-\delta_i{}^j& \frac{e\,\rho}3 \, \varepsilon_{ijk}\, x^k
\end{matrix} \bigg)
\eea
where $x^a=(x^i,\bar\pi_i)$ with $i=1,2,3$ and $a=1,\dots,6$. The Jacobiator is given by
\bea
\Pi^{abc} =\{x^a,x^b,x^c\}=
\bigg(\begin{matrix}
0&0\\
0&\frac{e\,\rho}3 \, \varepsilon_{ijk}
\end{matrix} \bigg)
\ .
\eea
Deformation quantisation of the twisted Poisson structure is determined by
 the bidifferential operator
\bea
\FF=\exp\big(-\mbox{$\frac{\ii\hbar}2$}\, \theta^{ab}(x)\, \partial_a\otimes\partial_b\big) \ ,
\eea
where $\partial_a=\frac\partial{\partial x^a}$, which defines a star product
\bea
f\star g:= \mbf\cdot\, \FF\cdot (f\otimes g)
\eea
that is a noncommutative and nonassociative deformation of the
pointwise product $f\, \mbf\cdot\, g$ of smooth functions $f,g\in C^\infty(\CP)$. Various useful properties of this nonassociative star product can be
found in~\cite{Mylonas2013,Kupriyanov2015}.

For constant monopole density, nonassociativity is controlled by the
multiplicative associator
\bea
f\star (g\star h)=\Phi\big((f\star g)\star h \big):= \star\, \exp\big(-\mbox{$\frac{\hbar^2}2$}\,
\Pi^{abc}\, \partial_a\otimes\partial_b\otimes\partial_c\big)\big((f\otimes
g)\otimes h\big) \ .
\label{eq:associator}\eea
The twisted coproduct of a vector field $X$ on $\CP$ is given
by $\Delta(X)=\FF\, (X\otimes1+1\otimes X)\,
\FF^{-1}$~\cite{Mylonas2013}; it determines the deformed Leibniz rule
\bea
X(f\star g) = \star\,\Delta(X)(f\otimes g) \ .
\eea
In particular, the twisted coproduct of primitive
translation generators is given by~\cite{Mylonas2013}
\bea
\Delta(\partial_a)=\partial_a\otimes1+1\otimes\partial_a+
\mbox{$\frac{\ii\hbar} 2$}\, \Pi_a{}^{bc}\, \partial_b\otimes\partial_c
\eea
which yields the deformed Leibniz rule
\bea
\partial_a(f\star g)= (\partial_af)\star
g+f\star(\partial_ag)+\mbox{$\frac{\ii\hbar}2$}\,
\Pi_a{}^{bc}\,( \partial_bf)\star(\partial_cg) \ .
\label{eq:Leibniz}\eea

\bigskip

\noindent
{\bf Composition product. \ } 
Recall the \emph{composition product} $\circ$ from~\cite{Mylonas2013}:
For functions $f,g,\varphi\in C^\infty(\CP)$, we define
\bea
(f\circ g)\star\varphi := f\star (g\star\varphi) \ .
\label{eq:circdefs}\eea
This defines a noncommutative product which is
\emph{associative} by construction, since by induction we have
\bea
(f_1\circ f_2\circ \cdots\circ f_n)\star\varphi =
f_1\star\big(f_2\star\big(\cdots\star (f_n\star\varphi)\cdots\big) \big) \ .
\eea
There is further a conjugate composition product $f\cirp g$ with the
property $(f\circ g)^*=g^*\cirp f^*$~\cite{Mylonas2013}, but we shall not need it
here. The
associativity properties of the star product $\star$ are completely
characterised by the composition products, in the sense that $\star$
is nonassociative if and only if there exist functions $f,g\in C^\infty(\CP)$
such that $f\circ g\notin C^\infty(\CP)$, 
while noncommutativity of
the compositions themselves are characterised by the commutators
$[f,g]_\circ:= f\circ g-g\circ f$. However, not all functions
need have this property; for example, in the case of a constant
monopole distribution $x^a\circ x^a=x^a\star x^a=x_a^2$;
see~\cite{Mylonas2013} for details.

For constant monopole density we can explicitly characterise the subalgebra
of differential operators ${\sf Diff}(\CP)$ on which the composition
products close in terms of the star product $\star$. For this, we use the definition \eqref{eq:circdefs}
and the associator relation \eqref{eq:associator} for arbitrary 
test
functions $\varphi\in C^\infty(\CP)$ to find
\bea
f\circ g &=& f\star g \label{eq:fcircg} \\ && +\, \sum_{n=1}^\infty\, \frac1{n!}\, \Big(
-\frac{\hbar^2}2\, \Big)^n\,
\Pi^{a_1b_1c_1}\cdots\Pi^{a_nb_nc_n}\,
\big((\partial_{a_1}\cdots\partial_{a_n}f)
\star(\partial_{b_1}\cdots\partial_{b_n}g)\big)\star \partial_{c_1}\cdots\partial_{c_n}
\nn
\eea
with $\partial_a\star\varphi:=\partial_a\varphi$. In particular, for
the coordinate generators we find
\bea
[x^a,x^b]_\circ = \ii\hbar\,
\theta^{ab}(x)-\hbar^2\,\Pi^{abc}\, \partial_c \ .
\eea
From the deformed Leibniz rule \eqref{eq:Leibniz} and the definition
$(\partial_a\circ
f)\star\varphi= \partial_a\star(f\star\varphi):=\partial_a(f\star
\varphi)$ we have
\bea
\partial_a\circ f
= \partial_af+f\star\partial_a+\mbox{$\frac{\ii\hbar}2$}\,
\Pi_a{}^{bc}\, (\partial_bf)\star \partial_c \ .
\eea

The general relations in ${\sf Diff}(\CP)$ can be obtained as follows. 
We can extend the coproduct $\Delta$ to arbitrary differential operators $D=\sum_k\,
d_{a_1\cdots a_k}(x) \, \partial_{a_1}\cdots\partial_{a_k}$ as an
algebra homomorphism with $\Delta(1)=1\otimes1$ and use the usual Sweedler notation (with implicit summation)
\bea
\Delta(D):= D_{(1)}\otimes D_{(2)} \ .
\eea
This encodes the deformed action of $D$ on star products $f\star
\varphi$ and for any $f\in C^\infty(\CP)$ we have
\bea
D\circ f= (D_{(1)}\cdot f)\star D_{(2)} \ ,
\eea
while $f\circ D$ is given by the formula \eqref{eq:fcircg} with $g$ replaced by $D$ and the derivatives $\partial_{b_i}$ acting on $d_{a_1\cdots a_k}$. Similarly, $f\circ\partial_b=f\star\partial_b$ and $\partial_a\circ\partial_b=\partial_a\star\partial_b$ so that
\bea
[\partial_a,\partial_b]_\circ=[\partial_a,\partial_b]_\star=0 \ .
\eea
In particular, the usual adjoint action of derivatives is modified by
nonassociativity as
\bea
[\partial_a,f]_\circ = \partial_af+\mbox{$\frac{\ii\hbar}2$}\,
\Pi_a{}^{bc}\,(\partial_bf)\star\partial_c \ . 
\eea
By a similar calculation we find
\bea
D_1\circ D_2= (D_{1\,(1)}\cdot D_2)\star D_{1\,(2)}
\eea
for any two differential operators $D_1,D_2\in\Diff(\CP)$.

In summary, starting from the nonassociative algebra
$(C^\infty(\CP),\star)$ we have explicitly constructed an associative
algebra $(\Diff(\CP),\circ)$ in which $C^\infty(\CP)$ is contained as a subspace (but not as a subalgebra). Notice that, in constrast to the $\star$-commutator, the $\circ$-commutator is a $\circ$-derivation,
\bea
[f,g\circ h]_\circ=[f,g]_\circ\circ h+g\circ[f,h]_\circ \ ,
\eea
and it satisfies the Jacobi identity by virtue of the associativity of
the composition product $\circ$. It is this feature which allows for
a consistent formulation of quantum dynamics in the Heisenberg picture
of nonassociative quantum mechanics: For a given Hamiltonian $\mcH$ and
an observable (real function) $\alg$ on
phase space $\CP$, time evolution can be defined with the $\circ$-commutator as
\bea
\frac{\dd \alg}{\dd t} = \frac\ii\hbar\, [\mcH,\alg]_\circ
\eea
since then the Leibniz rule consistently implies
\bea
\frac{\dd(\alg\circ \balg)}{\dd t}=\frac{\dd \alg}{\dd t}\circ \balg+\alg\circ\frac{\dd \balg}{\dd t} \ .
\eea
For example we have 
\bea
[x_i^2,\alg]_\circ = x^i\circ[x^i,\alg]_\circ+[x^i,\alg]_\circ\circ x^i = 
-2\ii\hbar\, x^i \, \frac{\partial \alg}{\partial \bar\pi_i} 
\eea
since $[x^i,\alg]_\circ=-\ii\hbar\, \frac{\partial \alg}{\partial \bar\pi_i}$.

\bigskip

\noindent
{\bf Symplectic realisation. \ } 
It is now relatively straightforward to see that the quantisation of
the symplectic algebra \eqref{PB1} agrees exactly with
the quantum $\circ$-brackets above. By writing $\tilde
x_a=(\tilde\pi_i,\tilde x^i)$ and using the generalised Bopp shift
of Appendix~\ref{app:symplecticPoisson}, these Poisson
brackets are quantised by the representation on the original algebra
of functions $C^\infty(\CP)$ given via the differential operators
\bea
\widehat{x}^{\,a}=x^a-\mbox{$\frac{\ii\hbar}2$}\, \theta^{ab}(x)\, \partial_b
\qquad \mbox{and} \qquad \widehat{\tilde x}_a=\ii\hbar\, \partial_a
\label{eq:diffBopp}\eea
which satisfy the non-vanishing commutation relations
\bea
[\,\widehat{x}^{\,a},\widehat{x}^{\,b}\,]=\ii\hbar\, \big(\theta^{ab} +
\Pi^{abc}\,\widehat{\tilde x}_c \big) \qquad \mbox{and} \qquad [\,\widehat{x}^{\,a},\widehat{\tilde x}_b\,]=\ii\hbar\,\big( \delta^a{}_b+\mbox{$\frac{1}2$}\,
(\partial_b\theta^{ac}) \,
\widehat{\tilde x}_c \big) \ .
\eea
Hence they reproduce the associative $\circ$-algebra of differential
operators, and in particular
\bea
f(\,\widehat{x}\,)g(x)=(f\circ g)(x)
\eea
for functions $f,g\in C^\infty(\CP)$. Moreover, as
already noted in~\cite{Mylonas2013,Kupriyanov2015}, these Bopp shifts
enable a rewriting of the nonassociative star product on functions
$f,g\in C^\infty(\CP)$ with integrable Fourier transforms as
\bea
(f\star g)(x) = f(\,\widehat{x}\,)\cdot g(x) =\int \, \dd w \  \tilde
f(w) \ \e^{\ii w_a\,\widehat{x}^{\,a}}\cdot g(x) \ ,
\eea
where 
\bea
\tilde f(w)=\frac1{(2\pi)^6}\, \int\, \dd x \ f(x) \ \e^{-\ii
  w_a\, x^a} \ .
\eea
Explicitly, the corresponding twisted Bopp shifts
\bea
\widehat{x}^{\,i}=x^i-\mbox{$\frac{\ii\hbar}2\, \frac\partial{\partial \bar\pi_i}$}
\qquad \mbox{and} \qquad \widehat{\bar\pi}_i=\bar\pi_i+\mbox{$\frac{\ii\hbar}2\, \big(\,
\frac\partial{\partial x^i}+ \frac{e\,\rho}3\, \varepsilon_{ijk}\, x^k\,
\frac\partial{\partial \bar\pi_j}\, \big) $}
\label{eq:HtwistedBopp}\eea
satisfy the non-vanishing commutation relations
\bea
\big[\widehat{x}^{\,i},\widehat{\bar\pi}_j \big]&=&
\big[\widehat{x}^{\,i},\widehat{\tilde\pi}_j \big] \ \ = \ \
\big[\,\widehat{\bar\pi}_i,\widehat{\tilde x}{}^{\,j} \big] \ \ = \ \ \ii\hbar\, \delta^i{}_j \ , \nn \\[4pt]
\big[\,\widehat{\bar\pi}_i,\widehat{\bar\pi}_j \big]&=& \mbox{$\frac{\ii\hbar\,e\,\rho}3$}\, \varepsilon_{ijk}\, \big(\widehat{x}^{\,k}-\widehat{\tilde x}{}^{\,k}\big) \ , \nn \\[4pt]
\big[\,\widehat{\bar\pi}_i,\widehat{\tilde\pi}_j \big]&=&
\big[\,\widehat{\tilde\pi}_i,\widehat{\bar\pi}_j \big] \ \ = \ \ \mbox{$\frac{\ii\hbar\,e\,\rho}6$}\, \varepsilon_{ijk}\, \widehat{\tilde x}{}^{\,k}
\ ,
\eea
with $\widehat{\tilde\pi}_i=\ii\hbar\, \frac\partial{\partial x^i}$ and $\widehat{\tilde x}{}^{\,i}=\ii\hbar\, \frac{\partial}{\partial \bar\pi_i}$. Written in this form, the symplectic algebra is similar to the Lie algebras of observables in geometric quantisation of twisted Poisson manifolds~\cite{Petalidou2007}. For $\rho=0$ the
differential operators \eqref{eq:HtwistedBopp} reduce to the usual
Bopp shifts of phase space quantum mechanics~\cite{Zachos2001}. 

For completeness, we note that the composition products 
also follow from deformation quantisation of the symplectic algebra
\eqref{PB1} itself by similarly defining a star product
\bea
(f\,\tilde\star\, g)(x,\tilde x) := f(\widehat{x},\widehat{\tilde x}\,)\cdot g(x,\tilde x)
\eea
on functions $f,g\in C^\infty(\CCP)$.
For this, let us rewrite the classical Poisson brackets in terms of a bigger algebra as
\bea
\{x^{\tilde a},x^{\tilde b}\}= \xi^{\tilde a \tilde b}+C^{\tilde a\tilde b}{}_{\tilde c}\, x^{\tilde c} \ ,
\eea
where $x^{\tilde a}=(x^a,\tilde x_a)$ with $\tilde a=1,\dots,12$, the non-vanishing central elements are
\bea
\xi^a{}_b=\delta^a{}_b \qquad \mbox{and} \qquad \xi^{x^i,\bar\pi_j}=\delta^i{}_j \ ,
\eea
and the non-vanishing structure constants are
\bea
C^{ab}{}_c= \Pi^{ab}{}_c \ , \qquad C^{abc}= -\Pi^{abc} \qquad \mbox{and} \qquad C^a{}_b{}^c=-\mbox{$\frac12$}\, \Pi^a{}_b{}^c \ .
\eea
We further rewrite these relations in the Lie algebraic form
\bea
\{\tilde x^{\tilde a},\tilde x^{\tilde b}\}=\tilde C^{\tilde a\tilde b}{}_{\tilde c}\, \tilde x^{\tilde c}
\eea
where $\tilde x^{\tilde a}=(x^{\tilde a},\kappa)$ with $\kappa$ central elements, and the non-vanishing structure constants are
\bea
\tilde C^{\tilde a\tilde b}{}_{ \kappa}=\xi^{\tilde a\tilde b} \qquad \mbox{and} \qquad \tilde C^{\tilde a\tilde b}{}_{\tilde c}=C^{\tilde a \tilde b}{}_{\tilde c} \ .
\eea
Now we can apply the polydifferential expansion
of~\cite[eq.~(3.4)]{Kupriyanov2015a} (which applies generally to
linear Poisson structures on $\R^d$) to functions $f$ which are
independent of $\kappa$, and after setting $\kappa=1$ we
arrive at the basic associative coordinate star products
\bea
x^{\tilde a}\, \tilde\star\, f&=& x^{\tilde a}\, \mbf\cdot\, f+\sum_{n=1}^\infty\, \frac{(-\ii\hbar)^n\, B_n}{n!} \, C^{\tilde a\tilde b_1}{}_{\tilde c_1}\, C^{\tilde c_1\tilde b_2}{}_{\tilde c_2}\cdots C^{\tilde c_{n-2}\tilde b_{n-1}}{}_{\tilde c_{n-1}}\, \nn\\ && \qquad \qquad \qquad\qquad \qquad\qquad \times \ \big(\xi^{\tilde c_{n-1}\tilde b_n}+C^{\tilde c_{n-1}\tilde b_n}{}_{\tilde c_n}\, x^{\tilde c_n}\big)\, \partial_{\tilde b_1}\cdots\partial_{\tilde b_n}f \ ,
\label{eq:xhatstarf}\eea
with $C^{\tilde c_{-2}\tilde b_{-1}}{}_{\tilde c_{-1}}:=1$, where $B_n$ are the Bernoulli numbers.

\subsection*{Acknowledgments}

We thank Chris Hull, Olaf Lechtenfeld, Dieter L\"ust, Emanuel Malek, Eric Plauschinn and Alexander Popov for helpful discussions.
This work was initiated
while R.J.S. was visiting the Centro de Matem\'atica, Computa\c{c}\~{a}o e
Cogni\c{c}\~{a}o of the Universidade de Federal do ABC in S\~ao Paulo,
  Brazil during June--July 2016, whom he warmly thanks for support and
  hospitality during his stay there. The authors acknowledge support from the Action MP1405 QSPACE from 
the European Cooperation in Science and Technology (COST), by the Consolidated Grant ST/P000363/1 
from the UK Science and Technology Facilities Council (STFC), by
the Visiting Researcher Program
Grant 2016/04341-5 from the Funda\c{c}\~{a}o de Amparo \'a Pesquisa do
Estado de S\~ao Paulo (FAPESP, Brazil), the Grant 305372/2016-5 from the Conselho Nacional de Pesquisa (CNPq, Brazil), and the Capes-Humboldt Fellowship 0079/16-2. 

\appendix

\newsection{Symplectic realisations of quasi-Poisson structures\label{app:symplecticPoisson}}

A \emph{symplectic realisation} of a Poisson structure $\theta$ on a
manifold $M$ is a symplectic manifold $(S,\Omega)$ together with a
surjective submersion ${\sf p}:S\to M$ which preserves the Poisson
structures: ${\sf p}_*\,\Omega^{-1} = \theta$. It is a fundamental
result in Poisson geometry that any Poisson manifold admits a
symplectic realisation. The original local construction for
$M=\real^d$ is due to~\cite{Weinstein-local}; it proceeds by taking
$S=T^*M$ to be the phase space of $M$, with the canonical projection
${\sf p}:T^*M\to M$, and $\Omega$ to be the integrated pullback of the
canonical symplectic structure $\dd p_i\wedge\dd x^i$ on $T^*M$ by the flow of the vector field $\theta^{ij}(x)\, p_i\, \partial_j$, where $(x,p)\in T^*M=\real^d\times(\real^d)^*$. The early global constructions based on integrating symplectic groupoids are due to~\cite{Karasev,Weinstein-groupoid,Weinstein}. The extension to almost symplectic realisations of twisted Poisson structures is established globally by~\cite{CattaneoXu}, while local symplectic realisations of arbitrary quasi-Poisson structures are constructed by~\cite{Kup14,SR}.

Given an arbitrary bivector $\Theta=\frac12\, \Theta^{ij}(x) \, \partial_i\wedge\partial_j$ on a manifold $M$ of dimension $d$, the algebra of \emph{quasi-Poisson brackets}
\begin{equation}
\{x^i,x^j\}=\alpha\, \Theta^{ij}(x) \ ,
\end{equation}
for local coordinates $x\in \real^d$ and a deformation parameter $\alpha\in\real$, is bilinear and antisymmetric but does not necessarily satisfy the Jacobi identity. Let
\bea
\{x^i,x^j,x^k\}=\mbox{$\frac13$}\,
\big(\{x^i,\{x^j,x^k\}\}+\mbox{cyclic}\big) = \alpha^2 \, \Pi^{ijk}(x)
\eea
be the corresponding Jacobiator $\Pi=\frac1{3!}\, \Pi^{ijk}(x)\, \partial_i\wedge\partial_j\wedge\partial_k$, where
\bea
\Pi^{ijk} = \mbox{$\frac{1}3$} \, \big(
\Theta^{il}\, \partial_l\Theta^{jk}+\Theta^{kl}\, \partial_l\Theta^{ij}+\Theta^{jl}\, \partial_l\Theta^{ki}
\big) \ .
\eea
If the bivector $\Theta$ is non-degenerate, it is easy to check that
\bea
\Pi^{ijk} = \mbox{$\frac13$}\, \Theta^{im}\, \Theta^{in}\, \Theta^{kl}\, \tau_{mnl} \qquad \mbox{with} \quad \tau_{mnl}=\partial_m\Theta^{-1}_{nl} +\partial_l\Theta^{-1}_{mn}+\partial_n\Theta^{-1}_{lm} \ ,
\eea
and in this case $\Theta$ defines a \emph{twisted Poisson bracket} with twisting three-form $\tau=\dd\Theta^{-1}$ on $M$.

We can ``double'' the local space to $\real^{2d}$ with coordinates $\xi^\mu=(x^i,{\tilde x}_i)$ for $\mu=1,\dots,2d$ and construct a Poisson bracket
\begin{equation}
\{\xi^\mu,\xi^\nu\}= \Omega^{\mu\nu}(\xi)=\Omega_0^{\mu\nu}+\alpha\, \Omega_1^{\mu\nu}(\xi)+O(\alpha^2)
\end{equation}
as a formal power series in the parameter $\alpha$, where $\Omega_0^{\mu\nu}$ is the canonical symplectic matrix. The Poisson brackets of the original coordinate functions are then
\begin{equation}
\{x^i,x^j\} = \alpha\, \omega^{ij}(x,\tilde x) \qquad \mbox{with} \quad \omega^{ij}(x,0)=\Theta^{ij}(x) \ .
\end{equation}
In particular
\begin{equation}
 \omega^{ij}(x,\tilde x)=\Theta^{ij}(x)-\alpha\, \Pi^{ijk}(x)\, \tilde x_k +O(\alpha^2) \ .
\end{equation}
The expansion may be explicitly constructed by introducing local Darboux coordinates $\eta^\mu=(y^i,\pi_i)$ and writing the generalised Bopp shift
\begin{equation}
x^i=y^i-\mbox{$\frac\alpha2$}\, \Theta^{ij}(y)\, \pi_j+O(\alpha^2) \qquad \mbox{and} \qquad \tilde x_i=\pi_i \ .
\end{equation}
See \cite{Kup14,SR} for further details of this construction. 

For the example of the magnetic monopole algebra \eqref{a1}, all
higher order corrections vanish and we arrive at the Poisson brackets 
\bea
\{x^i,x^j\} &=& \alpha\, \Theta^{ij}(x)-\alpha^2\, \Pi^{ijk}\, \tilde
x_k \ , \nn \\[4pt]
\{x^i,\tilde x_j\} &=& \delta^i{}_j + \mbox{$\frac\alpha2$}\,
(\partial_j\Theta^{ik})\, \tilde x_k \ , \nn \\[4pt]
\{\tilde x_i,\tilde x_j\} &=& 0 \ ,
\eea
which for $\alpha=1$ coincide with \eqref{PB0} and \eqref{PB}.

\newsection{Doubled harmonic oscillators\label{app:dho}}

The idea of employing additional degrees of freedom for the construction of variational principles for non-Lagrangian equations of motion appeared for the first time in the context of the one-dimensional damped harmonic oscillator with mass $m$, angular frequency $\omega$ and friction parameter $\lambda$, described by the equation of motion
\bea
m\, \ddot x+ \lambda\, \dot x+\omega^2\, x =0\ .\label{dho}
\eea
The Lagrangian is postulated to be the product of the original equation of motion and a Lagrange multiplier $\tilde x$, in the form $L_{\rm dho}=\tilde x\, \big(m\, \ddot x+\lambda\, \dot x+\omega^2\, x\big)$, see~\cite{Bateman}. Clearly the variation of this Lagrangian with respect to $\tilde x$ yields the original equation of motion (\ref{dho}). However, the variation of $L_{\rm dho}$ with respect to $x$ gives the equation of motion for an additional ``double'' oscillator
\bea
m\, \ddot{\tilde x}- \lambda\, \dot {\tilde x}+\omega^2\, \tilde x=0\ ,\label{dho2}
\eea
which is the time-reversed image of the original oscillator in the sense that $\lambda\to-\lambda$. 
The total energy of the doubled system is conserved meaning that the energy dissipated by the first oscillator \eqref{dho} is absorbed by the second oscillator (\ref{dho2}) which thereby plays the role of an effective reservoir.

For $\lambda=0$ the Hamiltonian reads
\bea
\mcH_{\rm dho}=\mbox{$\frac 1m$}\, p\, \tilde p+\omega^2\, x\, \tilde x\ , \label{do1}
\eea
with $p=m\,\dot x$, $\tilde p=m\, \dot{\tilde x}$ and the canonical Poisson brackets between all variables. It can be represented as
\bea
\mcH_{\rm dho}=\mcH_+-\mcH_-:=\big(\mbox{$\frac{1}{2m}$}\, p_+^2+\mbox{$\frac{\omega^2}{2}$}\, x_+^2\big) - \big(\mbox{$\frac{1}{2m}$}\, p_-^2+\mbox{$\frac{\omega^2}{2}$}\, x_-^2\big) \ ,\label{do3}
\eea
where
\bea
 p_\pm=\mbox{$\frac{1}{\sqrt 2}$}\, \big(p\pm\tilde p\big)\qquad \mbox{and} \qquad x_\pm =
 \mbox{$\frac{1}{\sqrt 2}$}\, \big(x\pm\tilde x\big) \label{do5}
\eea
defines a canonical transformation of phase space coordinates. The Hamiltonian $\mcH_{\rm dho}$ is not positive and does not represent the energy of the doubled oscillator system. The energy of each subsystem is defined by $\mcH_\pm$ respectively, while the total energy is determined as $E_{\rm dho}=\mcH_++\mcH_-$.

The quantisation of this model was discussed in the context of quantum dissipation, see e.g.~\cite{Feshbach,Dekker,Celeghini,tHooft} for early works. For this, we introduce creation and annihilation operators by
\bea
\widehat{a}_\pm = \sqrt{\mbox{$\frac{m\,\omega}{2\hbar}$}} \, \big(\,\widehat{ x}_\pm+
\mbox{$\frac{\ii}{m\,\omega}$}\, \widehat{p}_\pm \big)\qquad
\mbox{and} \qquad {\widehat a}{}^\dagger_\pm =
\sqrt{\mbox{$\frac{m\,\omega}{2\hbar}$}} \, \big(\,\widehat{ x}_\pm- \mbox{$\frac{\ii}{m\,\omega}$}\, \widehat{p}_\pm \big) \ .\label{do6}
\eea
One easily checks
\bea
\big[ \widehat{a}_\pm\,,\,\widehat{a}{}^\dag_\pm\big]=1\ ,\label{do7}
\eea
while all other commutators vanish. In terms of the creation and
annihilation operators (\ref{do6}), the quantum Hamiltonian
corresponding to (\ref{do3}) becomes
\bea
\widehat{\mcH}_{\rm dho}=\widehat{\mcH}_+-\widehat{\mcH}_-:= {\hbar\,\omega}\, \big(\,\widehat{a}{}^\dag_+ \, \widehat{a}_+-\widehat{a }{}^\dag_- \, \widehat{a}_--1\big)\ .\label{do8}
\eea
The Hamiltonian \eqref{do8} is unbounded, but it can be expressed as
the difference of two Hamiltonians $\widehat{\mcH}_\pm$
of harmonic oscillator type, which are bounded from below and commute,
$\big[ \widehat{\mcH}_+, \widehat{\mcH}_-\big]=0$. The Hamiltonian
$\widehat{\mcH}_{\rm dho}$ can be represented on the two-particle bosonic Fock space $\CCF$. In the standard number basis, its eigenstates are given by 
\bea
|n_+,n_-\rangle=\frac{\big(\,\widehat{a}{}^\dag_+\big)^{n_+}}{\sqrt{n_+!}}\,\frac{\big(\,\widehat{a}{}^\dag_-\big)^{n_-}}{\sqrt{n_-!}}|0,0\rangle
\eea
for $n_\pm\in\N_0$ with corresponding eigenvalues
$\hbar\,\omega\,(n_+-n_--1)$, where $\widehat{a}{}^\dag_\pm \,
\widehat{a}_\pm |n_+,n_-\rangle=n_\pm\,|n_+,n_-\rangle$ with
$\widehat{a}_+|0,n_-\rangle=0=\widehat{a}_-|n_+,0\rangle$. In
particular, the usual harmonic oscillator eigenstates and eigenvalues
emerge only when one sets the $\widehat{a}_-$-oscillator in its ground state
$|n_+,0\rangle$, thereby turning off the reservoir which is needed
only when $\lambda\neq0$ in order to absorb the energy dissipated by
the physical $\widehat{a}_+$-oscillator.

\newsection{Magnetic duality and locally non-geometric fluxes\label{app:nongeometric}}

Our constructions can also have ramifications for the phase space
structures of locally non-geometric fluxes in string theory and
M-theory. For this, let us write the symplectic algebra
\eqref{a5} as
\bea
\{x^i,\pi_j\}&=&\{\tilde x^i,\tilde\pi_j\} \ \ = \ \ \delta^i{}_j \ , \nn\\[4pt]
\{\pi_i,\pi_j\}&=& H_{ijk} \, \tilde x^k\ , \nn\\[4pt]
\{\pi_i,\tilde\pi_j\}&=&\{\tilde\pi_i,\pi_j\} \ \ = \ \ -\mbox{$\frac 12$}\, H_{ijk}\, x^k \ ,\label{a5d}
\eea
where generally $H=\frac1{3!}\, H_{ijk}\, \dd x^i\wedge\dd x^j\wedge\dd x^k$ is
a constant three-form on $\real^d$; for the $d=3$ example of the spherically
symmetric magnetic field \eqref{eq:Brot} considered in the main text,
we take
$H_{ijk}=-\frac{e\,\rho}3\,\varepsilon_{ijk}$. Now we can adapt the
magnetic duality transformation of order four from~\cite{Lust:2017bgx}
to our situation to get
\bea
(x^i,\tilde x^i)\longmapsto (-\pi_i,-\tilde\pi_i) \ , \qquad
(\pi_i,\tilde\pi_i)\longmapsto (x^i,\tilde x^i) \qquad \mbox{and}
\qquad H_{ijk}\longmapsto -R^{ijk} \ ,
\eea
where $R=\frac1{3!}\, R^{ijk}\, \partial_i\wedge\partial_j\wedge \partial_k$ is
a constant trivector on $\real^d$; for $d=3$ it corresponds to a
locally non-geometric flux $R^{ijk}=\frac{\ell_s^3}{3\hbar^2}\, N\,
\varepsilon^{ijk}$ in IIA string theory with $N$ units of NS--NS flux,
where $\ell_s$ is the string length. Under this map the Poisson
brackets \eqref{a5d} become
\bea
\{x^i,\pi_j\}&=&\{\tilde x^i,\tilde\pi_j\} \ \ = \ \ \delta^i{}_j \ , \nn\\[4pt]
\{x^i,x^j\}&=& R^{ijk} \, \tilde\pi_k \ , \nn\\[4pt]
\{x^i,\tilde x^j\}&=&\{\tilde x^i,x^j\} \ \ = \ \ -\mbox{$\frac 12$}\, R^{ijk}\, \pi_k \ ,\label{a5dual}
\eea
which we identify as the symplectic realisation of the nonassociative
phase space algebra of closed strings propagating in locally
non-geometric flux backgrounds~\cite{Lust2010,Mylonas2012}.

It is tempting in this setting to compare our symplectic realisation
with the perspective of double field theory, wherein auxiliary winding
coordinates are introduced in order to construct a theory with manifest
$O(d,d)$ symmetry (see e.g.~\cite{dftrev1,dftrev2,dftrev3} for reviews). In this case, only after eliminating the dependence
of fields on the winding coordinates, or more generally by choosing a polarisation
which halves the number of extended space coordinates (by weak or strong
constraints), does one speak of ``physical'' coordinates. However, in
our case we have seen that there is no possibility to choose different such
``polarisations'' to get to a physical space with a nonassociative
algebra that can be obtained from reduction of our fixed symplectic
algebra. Furthermore, this symplectic algebra is very different from
the double field theory phase space model of~\cite{Lust2010,Blumenhagen2011,BL}, which
still involves a nonassociative algebra, whereas the complete algebra
\eqref{a5dual} is associative.

Our symplectic realisation is also useful for further investigation of
the nonassociative phase space algebra of M2-branes propagating in
four-dimensional locally non-geometric flux backgrounds of
M-theory~\cite{Gunaydin2016,Kupriyanov2017}.\footnote{In this setting the
``quasi-nonassociative'' quantum mechanics of~\cite{Aizawa:2017nue}
may be relevant for understanding the interplay between associative
and nonassociative structures.} The magnetic dual of this
configuration was identified by~\cite{Lust:2017bgx} as the phase space
algebra of a non-geometric Kaluza-Klein monopole in M-theory. As the
symplectic realisation provides an explicit globally-defined magnetic
vector potential $A_I(x^I)$, this may be used to construct this geometry
more precisely as an explicit supergravity solution.


\begin{thebibliography}{99}

\baselineskip=12pt

\bibitem{Aizawa:2017nue}
  N.~Aizawa, Z.~Kuznetsova and F.~Toppan,
  ``The quasi-nonassociative exceptional $F(4)$ deformed quantum oscillator,''
  J.\ Math.\ Phys.\  {\bf 59} (2018) 022101
  [arXiv:1711.02923 [math-ph]].

	\bibitem{dftrev1}
	G.~Aldazabal, D.~Marqu\'es and C.~N\'u\~nez,
	``Double field theory: A pedagogical review,''
	Class.\ Quant.\ Grav.\  {\bf 30} (2013) 163001
	[arXiv:1305.1907 [hep-th]].
	
\bibitem{BL}
I.~Bakas and D.~L\"ust,
  ``Three-cocycles, nonassociative star products and the magnetic paradigm of $R$-flux string vacua,''
  JHEP {\bf 1401} (2014) 171
  [arXiv:1309.3172 [hep-th]].
  
\bibitem{Batalin1986}
  I.A.~Batalin and E.S.~Fradkin,
  ``Operatorial quantization of dynamical systems subject to second class constraints,''
  Nucl.\ Phys.\ B {\bf 279} (1987) 514--528.
  
\bibitem{Batalin1989}
  I.A.~Batalin, E.S.~Fradkin and T.E.~Fradkina,
  ``Another version for operatorial quantization of dynamical systems with irreducible constraints,''
  Nucl.\ Phys.\ B {\bf 314} (1989) 158--174.
  
\bibitem{Bateman} 
H. Bateman, 
``On dissipative systems and related variational principles,''
Phys.\ Rev. {\bf 38} (1931) 815--819.

	\bibitem{dftrev2}
	D.S.~Berman and D.C.~Thompson,
	``Duality symmetric string and M-theory,''
	Phys.\ Rept.\  {\bf 566} (2014) 1--60
	[arXiv:1306.2643 [hep-th]].

\bibitem{Blumenhagen2010}
R.~Blumenhagen and E.~Plauschinn,
 ``Nonassociative gravity in string theory?,''
J.\ Phys.\ A {\bf 44} (2011) 015401
[arXiv:1010.1263 [hep-th]].

\bibitem{Blumenhagen2011}
R.~Blumenhagen, A.~Deser, D.~L\"ust, E.~Plauschinn and F.~Rennecke,
 ``Non-geometric fluxes, asymmetric strings and nonassociative geometry,''
J.\ Phys.\ A {\bf 44} (2011) 385401
[arXiv:1106.0316 [hep-th]].
	
\bibitem{Blumenhagen2013}
R.~Blumenhagen, M.~Fuchs, F.~Ha\ss ler, D.~L\"ust and R.~Sun,
``Nonassociative deformations of geometry in double field theory,''
JHEP {\bf 1404} (2014) 141 
[arXiv:1312.0719 [hep-th]].
	
\bibitem{Bojowald2014}
  M.~Bojowald, S.~Brahma, U.~B\"uy\"uk\c{c}am and T.~Strobl,
  ``States in nonassociative quantum mechanics: Uncertainty relations and semiclassical evolution,''
  JHEP {\bf 1503} (2015) 093
  [arXiv:1411.3710 [hep-th]].

\bibitem{Bojowald2015}
  M.~Bojowald, S.~Brahma and U.~B\"uy\"uk\c{c}am,
  ``Testing nonassociative quantum mechanics,''
  Phys.\ Rev.\ Lett.\  {\bf 115} (2015) 220402
  [arXiv:1510.07559 [quant-ph]].

\bibitem{Bunkinprep}
S.~Bunk, L.~M\"uller and R.J.~Szabo,
``Geometry and 2-Hilbert space for nonassociative magnetic
translations,''
arXiv:1804.08953 [hep-th].

\bibitem{Cabibbo1962}
N.~Cabibbo and E.~Ferrari,
``Quantum electrodynamics with Dirac monopoles,''
Nuovo Cimento {\bf 23} (1962) 1147--1154.

\bibitem{Carinena:2009ug}
  J.F.~Cari\~{n}ena, J.M.~Gracia-Bond\'{\i}a, F.~Lizzi, G.~Marmo and P.~Vitale,
  ``Star product in the presence of a monopole,''
  Phys.\ Lett.\ A {\bf 374} (2010) 3614--3618
  [arXiv:0912.2197 [math-ph]].

\bibitem{CattaneoXu}
A.S. Cattaneo and P. Xu, 
``Integration of twisted Poisson structures,'' 
J. Geom. Phys. {\bf 49} (2004) 187--196
[arXiv:math.SG/0302268].

\bibitem{Celeghini} 
  E.~Celeghini, M.~Rasetti and G.~Vitiello,
  ``Quantum dissipation,''
  Ann. Phys.\ {\bf 215} (1992) 156--170.

 \bibitem{Weinstein} 
 A. Coste, P. Dazord and A. Weinstein, 
 ``Groupoides symplectiques,'' 
 Publ. D\'ep. Math. Nouvelle S\'er. A {\bf 2} (1987) 1--62.
 
\bibitem{Dekker} 
H. Dekker, 
``Classical and quantum mechanics of the damped harmonic oscillator,'' 
Phys.\ Rept. {\bf 80} (1981) 1--112.

\bibitem{Dirac1931}
  P.A.M.~Dirac,
  ``Quantised singularities in the electromagnetic field,''
  Proc.\ Roy.\ Soc.\ London\ A {\bf 133} (1931) 60--72.

\bibitem{Dirac1948}
  P.A.M.~Dirac,
  ``The theory of magnetic poles,''
  Phys.\ Rev.\  {\bf 74} (1948) 817--830.
  
\bibitem{Faddeev1984}
  L.D.~Faddeev,
  ``Operator anomaly for the Gauss law,''
  Phys.\ Lett.\ B {\bf 145} (1984) 81--84.
  
\bibitem{Faddeev1986}
  L.D.~Faddeev and S.L.~Shatashvili,
  ``Realization of the Schwinger term in the Gauss law and the possibility of correct quantization of a theory with anomalies,''
  Phys.\ Lett.\ B {\bf 167} (1986) 225--228.
  
\bibitem{Feinberg1991}
  J.~Feinberg and M.~Moshe,
  ``Faddeev-Batalin-Fradkin extended phase space for superparticle quantization,''
  Ann. Phys.\  {\bf 206} (1991) 272--317.
  
\bibitem{Feshbach}
H. Feshbach and Y. Tikochinsky, 
``Quantisation of the damped harmonic oscillator,''  
Trans. New York Acad. Sci. {\bf 38} (1977) 44--53.

\bibitem{Gunaydin2016}
M.~G{\"u}naydin, D.~L{\"u}st and E.~Malek, 
``Nonassociativity in
  non-geometric string and M-theory backgrounds, the algebra of octonions, and
  missing momentum modes,''
JHEP {\bf 1611} (2016) 027
  [arXiv:1607.06474 [hep-th]].

\bibitem{Hannabuss2017}
  K.C.~Hannabuss,
  ``T-duality and the bulk-boundary correspondence,''
  J.\ Geom.\ Phys.\ {\bf 124} (2018) 421--435
  [arXiv:1704.00278 [cond-mat.mes-hall]].

	\bibitem{dftrev3}
	O.~Hohm, D.~L\"ust and B.~Zwiebach,
	``The spacetime of double field theory: Review, remarks, and outlook,''
	Fortsch.\ Phys.\  {\bf 61} (2013) 926--966
	[arXiv:1309.2977 [hep-th]].

 \bibitem{tHooft}
  G.~'t Hooft,
  ``Determinism and dissipation in quantum gravity,''
  Subnucl.\ Ser.\ {\bf 37} (2001) 397--430
  [arXiv:hep-th/0003005].

\bibitem{Jackiw1985}
R.~Jackiw, 
``Three-cocycle in mathematics and physics,'' 
Phys. Rev. Lett. {\bf 54} (1985) 159--162.

\bibitem{Karasev} 
M.V. Karasev, 
``Analogues of objects of the theory of Lie groups for nonlinear Poisson brackets,'' 
Math. USSR-Izv. {\bf 28} (1987) 497--527.

\bibitem{Kup14}
  V.G.~Kupriyanov,
  ``Quantum mechanics with coordinate dependent noncommutativity,''
  J.\ Math.\ Phys.\ {\bf 54} (2013) 112105
  [arXiv:1204.4823 [math-ph]].
  
 \bibitem{SR} 
 V.G.~Kupriyanov,
  ``Recurrence relations for symplectic realization of (quasi-)Poisson structures,''
  arXiv:1805.12040 [math-ph].

\bibitem{Kupriyanov2017}
  V.G.~Kupriyanov and R.J.~Szabo,
  ``$G_{2}$-structures and quantisation of non-geometric M-theory backgrounds,''
  JHEP {\bf 1702} (2017) 099
  [arXiv:1701.02574 [hep-th]].

\bibitem{Kupriyanov2015}
  V.G.~Kupriyanov and D.V.~Vassilevich,
  ``Nonassociative Weyl star products,''
  JHEP {\bf 1509} (2015) 103
  [arXiv:1506.02329 [hep-th]].

\bibitem{Kupriyanov2015a}
  V.G.~Kupriyanov and P.~Vitale,
  ``Noncommutative $ \R^d $ via closed star product,''
  JHEP {\bf 1508} (2015) 024
  [arXiv:1502.06544 [hep-th]].

\bibitem{Lust2010}
  D.~L\"ust,
  ``T-duality and closed string noncommutative (doubled) geometry,''
  JHEP {\bf 1012} (2010) 084
  [arXiv:1010.1361 [hep-th]].
  
	\bibitem{Lust2012}
	D.~L\"ust,
	``Twisted Poisson structures and noncommutative/nonassociative closed string geometry,''
	PoS CORFU {\bf 2011} (2011) 086
	[arXiv:1205.0100 [hep-th]].

\bibitem{Lust:2017bgx}
  D.~L\"ust, E.~Malek and R.J.~Szabo,
  ``Non-geometric Kaluza-Klein monopoles and magnetic duals of M-theory $R$-flux backgrounds,''
  JHEP {\bf 1710} (2017) 144
  [arXiv:1705.09639 [hep-th]].
  
\bibitem{Mylonas2012}
  D.~Mylonas, P.~Schupp and R.J.~Szabo,
  ``Membrane sigma-models and quantisation of non-geometric flux backgrounds,''
  JHEP {\bf 1209} (2012) 012
  [arXiv:1207.0926 [hep-th]].

\bibitem{Mylonas2013}
  D.~Mylonas, P.~Schupp and R.J.~Szabo,
  ``Non-geometric fluxes, quasi-Hopf twist deformations and nonassociative quantum mechanics,''
  J.\ Math.\ Phys.\  {\bf 55} (2014) 122301
  [arXiv:1312.1621 [hep-th]].
  
\bibitem{Mylonas2014}
  D.~Mylonas, P.~Schupp and R.J.~Szabo,
  ``Nonassociative geometry and twist deformations in non-geometric string theory,''
  PoS ICMP {\bf 2013} (2013) 007
  [arXiv:1402.7306 [hep-th]].

\bibitem{Petalidou2007}
F.~Petalidou,
``On the geometric quantisation of twisted Poisson manifolds,''
J. Math. Phys. {\bf 48} (2007) 083502
[arXiv:0704.2989 [math.DG]].

\bibitem{Plyushchay2000}
  M.S.~Plyushchay,
  ``Monopole Chern-Simons term: Charge monopole system as a particle with spin,''
  Nucl.\ Phys.\ B {\bf 589} (2000) 413--439
  [arXiv:hep-th/0004032].

\bibitem{Plyushchay2013}
  M.S.~Plyushchay and A.~Wipf,
  ``Particle in a self-dual dyon background: Hidden free nature, and exotic superconformal symmetry,''
  Phys.\ Rev.\ D {\bf 89} (2014) 045017
  [arXiv:1311.2195 [hep-th]].

\bibitem{Soloviev2017}
  M.A.~Soloviev,
  ``Dirac's magnetic monopole and the Kontsevich star product,''
  J.\ Phys.\ A {\bf 51} (2018) 095205
  [arXiv:1708.05030 [math-ph]].

\bibitem{SzaboISQS}
R.J.~Szabo,
  ``Magnetic monopoles and nonassociative deformations of quantum theory,''
  J. Phys. Conf. Ser. {\bf 965} (2018) 012041
  [arXiv:1709.10080 [hep-th]].
  
\bibitem{Weinstein-local}
A. Weinstein, 
``The local structure of Poisson manifolds,'' 
J. Diff. Geom. {\bf 18} (1983) 523--557.

\bibitem{Weinstein-groupoid}
A. Weinstein, 
``Symplectic groupoids and Poisson manifolds,'' 
Bull. Amer. Math. Soc. {\bf 16} (1987) 101--104.

\bibitem{Wu1976}
  T.T.~Wu and C.N.~Yang,
  ``Dirac monopole without strings: Monopole harmonics,''
  Nucl.\ Phys.\ B {\bf 107} (1976) 365--380.

\bibitem{Zachos2001}
  C.K.~Zachos,
  ``Deformation quantisation: Quantum mechanics lives and works in phase space,''
  Int.\ J.\ Mod.\ Phys.\ A {\bf 17} (2002) 297--316
  [arXiv:hep-th/0110114].

\bibitem{Zak1964}
  J.~Zak,
  ``Magnetic translation group,''
  Phys.\ Rev.\  {\bf 134} (1964) A1602--A1606.

\bibitem{Zwanziger1968}
  D.~Zwanziger,
  ``Exactly soluble nonrelativistic model of particles with both electric and magnetic charges,''
  Phys.\ Rev.\  {\bf 176} (1968) 1480--1488.

\end{thebibliography}
\end{document}